\documentclass[apjl]{emulateapj}
\usepackage{apjfonts}
\usepackage{ulem}
\usepackage{verbatim}

\newcommand{\teff}{$T_{\!\mbox{\tiny\it eff}}$}

\newcommand{\hi}{H\,{\sc i}\rm}
\newcommand{\hii}{H\,{\sc ii}\rm}
\newcommand{\hiiit}{H\,{\footnotesize II}\rm}

\newcommand{\hei}{He\,{\sc i}\rm}
\newcommand{\heii}{He\,{\sc ii}\rm}
\newcommand{\siii}{[S\,{\sc iii}]}
\newcommand{\siv}{[S\,{\sc iv}]}
\newcommand{\nii}{[N\,{\sc ii}]}

\newcommand{\oiii}{[O\,{\sc iii}]}
\newcommand{\oii}{[O\,{\sc ii}]}
\newcommand{\oi}{[O\,{\sc i}]}
\newcommand{\oone}{O\,{\sc i}}
\newcommand{\mgi}{[Mg\,{\sc i}]}
\newcommand{\sii}{[S\,{\sc ii}]}
\newcommand{\ariii}{[Ar\,{\sc iii}]}

\newcommand{\neiii}{[Ne\,{\sc iii}]}
\newcommand{\neii}{[Ne\,{\sc ii}]}

\newcommand{\feiii}{[Fe\,{\sc iii}]}

\newcommand{\ciii}{C\,{\sc iii}\rm}
\newcommand{\niii}{N\,{\sc iii}\rm}
\newcommand{\civ}{C\,{\sc iv}\rm}
\newcommand{\nv}{N\,{\sc v}\rm}

\newcommand{\op}{O$^{+}$}
\newcommand{\hp}{H$^{+}$}
\newcommand{\np}{N$^{+}$}
\newcommand{\sulp}{S$^{+}$}
\newcommand{\opp}{O$^{++}$}
\newcommand{\spp}{S$^{++}$}
\newcommand{\nepp}{Ne$^{++}$}
\newcommand{\arpp}{Ar$^{++}$}

\newcommand{\te}{$T_e$}
\newcommand{\den}{$ N_e$}
\newcommand{\hgamma}{H$\gamma$}
\newcommand{\hbeta}{H$\beta$}
\newcommand{\halpha}{H$\alpha$}
\newcommand{\lin}{$\,\lambda$}
\newcommand{\llin}{$\,\lambda\lambda$}

\newcommand{\rtf}{$R_{25}$}

\newcommand{\oh}{12\,+\,log(O/H)}

\newcommand{\rtwothree}{R$_{23}$}
\newcommand{\vs}{vs.}

\slugcomment{}

\shorttitle{Abundances in NGC~300}
\shortauthors{Bresolin et al.}

\begin{document}

\title{Extragalactic chemical abundances: do \hii\ regions and young stars tell the same story? The case of the spiral galaxy NGC~300.\footnotemark[1]\\[3mm]} 

\footnotetext[1]{Based on observations collected at the European Southern Observatory, Chile, under program 077.B-0269.}

\author{Fabio Bresolin} \affil{Institute for Astronomy, 2680 Woodlawn Drive, Honolulu, HI 96822, USA\\ \vspace{-0.6cm}}
\author{Wolfgang Gieren} \affil{Universidad de Concepci\'on, Departamento de Astronomia, Casilla 160-C, Concepci\'on, Chile\\ \vspace{-0.6cm}}
\author{Rolf-Peter Kudritzki} \affil{Institute for Astronomy, 2680 Woodlawn Drive, Honolulu, HI 96822, USA\\ \vspace{-0.6cm}}
\author{Grzegorz Pietrzy\'nski} \affil{Universidad de Concepci\'on, Departamento de Astronomia, Casilla 160-C, Concepci\'on, Chile} \affil{Warsaw University Observatory Al Ujazdowskie 4, 00-478 Warsaw, Poland\\ \vspace{-0.6cm}}
\author{Miguel A. Urbaneja} \affil{Institute for Astronomy, 2680 Woodlawn Drive, Honolulu, HI 96822, USA\\ \vspace{-0.3cm}}
\and
\author{Giovanni Carraro} \affil{European Southern Observatory, Alonso de Cordova 3107, Santiago, Chile\\ \vspace{-0.cm}}

\begin{abstract}
We have obtained new spectrophotometric data for 28 \hii\/ regions in the spiral galaxy NGC~300, a member of the nearby Sculptor Group. The detection of several auroral lines, including \oiii\lin4363, \siii\lin6312 and \nii\lin5755, has allowed us to measure electron temperatures and direct chemical abundances for the whole sample.
We determine for the first time in this galaxy a radial gas-phase oxygen abundance gradient based solely on auroral lines, and obtain the following least-square solution: \oh\,=\,$8.57~(\pm 0.02) - 0.41~(\pm 0.03)~R/R_{25}$, where the galactocentric distance is expressed in terms of the isophotal radius \rtf. 
The characteristic oxygen abundance, measured at 0.4$\times$\rtf,  is \oh\,=\,8.41. 
The gradient corresponds to $-0.077\pm0.006$ dex\,kpc$^{-1}$, and agrees very well with the galactocentric trend in metallicity obtained for 29 B and A supergiants in the same galaxy, $-0.081\pm0.011$ dex\,kpc$^{-1}$.
The intercept of the regression for the nebular data virtually coincides with the 
intercept obtained from the stellar data, which is $8.59~(\pm 0.05)$.
This allows little room  for depletion of nebular oxygen onto dust grains, although in this kind of comparison we are somewhat limited by systematic uncertainties, such as those related to the atomic parameters used to derive the chemical compositions.

We discuss the implications of our result with regard to strong-line abundance indicators commonly used to estimate the chemical compositions of star-forming galaxies, such as \rtwothree. By applying a few popular calibrations of these indices based on grids of photoionization models on the NGC~300 \hii\/ region fluxes we find metallicities that are higher by 0.3 dex (a factor of two) or more relative to our nebular (\te-based) and stellar ones.

We detect Wolf-Rayet stellar emission features in $\sim$1/3 of our \hii\/ region spectra, and find that in one 
of the nebulae hosting these hot stars the ionizing field has a particularly hard spectrum, as gauged by the `softness' parameter $\eta$\,=\,(\op/\opp)/(\sulp/\spp). We suggest that this is related to the presence of an early WN star. 
By considering a larger sample of extragalactic \hii\/ regions we confirm, using direct abundance measurements, 
previous findings of a metallicity dependence of $\eta$, in the sense that softer stellar continua are found at high metallicity.

\end{abstract}

 
\section{Introduction}

The spectral analysis of \hii\/ regions has been an invaluable tool in astrophysics for the past few decades,  providing a straightforward means to measure present-day chemical abundances in a variety of galactic environments, which has led, for example, to the study of radial abundance gradients in spiral galaxies (\citealt{Vila-Costas:1992, Zaritsky:1994}).
It has recently become possible to extend the nebular techniques to star-forming galaxies at high redshift (\citealt{Pettini:2001, Shapley:2004}), allowing the investigation of cosmic chemical enrichment (\citealt{Kobulnicky:2004, Savaglio:2005, Maier:2006, Liu:2008, Cowie:2008})
in connection to fundamental properties of galaxies such as the mass-metallicity relation (\citealt{Lequeux:1979, Tremonti:2004, Erb:2006, Maiolino:2008, Perez-Montero:2008}).

The work on young massive stars to obtain reliable metallicities  in nearby galaxies is less mature, due to the need for a sophisticated NLTE treatment of the physical processes involving millions of metal lines in expanding atmospheres (\citealt{Hillier:1998, Pauldrach:2001, Puls:2005}), and the requirement for telescopes with large collecting areas to secure spectra of individual stars located a few Mpc away (\citealt{Bresolin:2001, Bresolin:2002a}). While it is possible to measure stellar metallicities  from the integrated spectra of young star clusters (\citealt{Larsen:2006, Larsen:2008}) or for star-forming galaxies at high redshift (\citealt{Rix:2004, Halliday:2008}), more stringent tests that compare the chemical compositions of galaxies as obtained from \hii\/ regions and massive stars should be carried out in nearby, well-resolved systems. In doing so, 
the comparison is limited to chemical elements that are measurable in both types of objects and  that, contrary to nitrogen, are largely unaffected by rotational mixing (\citealt{Maeder:2000, Hunter:2007}). For these reasons the abundances of oxygen, and more rarely iron, can be directly compared between ionized nebulae and young stars, mostly early-B dwarfs (within the Milky Way and in the Magellanic Clouds), and brighter A and B supergiants (in more distant galaxies). The expectation is that, once evolutionary effects in stars are properly accounted for, the present-day abundances derived from young massive stars and \hii\/ regions agree within the uncertainties of the measurements and of the modeling.

In low-metallicity and generally chemically homogeneous galaxies, such as the Magellanic Clouds and a small number of dwarf irregulars of the Local Group, the agreement found between \hii\/ region and young star chemical abundances is satisfactory (e.g.~\citealt{Trundle:2005, Bresolin:2006a, Lee:2006}). The possibilities for comparison are even fewer in the case of spiral galaxies, as only data for the Milky Way (\citealt{Rolleston:2000,Deharveng:2000,Daflon:2004})
and M33 (\citealt{Vilchez:1988,Urbaneja:2005a,Magrini:2007, Rosolowsky:2008}) have insofar allowed meaningful comparisons. Yet, these are perhaps the most interesting cases, because the metallicity in spiral galaxies can span a wide range, from the metal-rich nuclear regions to the metal-poor outskirts, of up to 1 dex (as in the case of M101, \citealt{Kennicutt:2003}).

A well-documented complication in \hii\/ region studies arises as the metallicity approaches the solar one [we adopt \oh$_\odot$\,=\,8.66 from \citealt{Asplund:2005}]. The classical method of nebular abundance analysis is based on the measurement of the \oiii\lin4363/\lin5007 line ratio, which is highly sensitive to the electron temperature of the gas, upon which the line emissivities strongly depend. As the nebular cooling shifts from optical to IR transitions with increasing metallicity (for example, approaching the centers of spiral galaxies), the auroral line \oiii\lin4363 becomes extremely faint, and virtually unobservable in extragalactic \hii\/ regions near the solar metallicity, except for the brightest objects.
The common solution is to base the abundance measurements on line ratios that involve the strongest collisionally excited lines that are present in nebular spectra, such as \rtwothree\,=\,(\oii\lin3727+\oiii\llin4959,5007)/H$\beta$ (\citealt{Pagel:1979}).
However, the use of these techniques is not exempt from difficulties, and in particular different calibrations of `strong-line indices' in terms of the oxygen abundance lead to poorly understood systematic discrepancies (\citealt{Bresolin:2008, Kewley:2008}). Alternatively, deep observations aimed at the detection of \oiii\lin4363 or other auroral lines, such as \siii\lin6312 and \nii\lin5575, can be carried out to by-pass these calibration issues, in order to have a `direct' measurement of the chemical abundances from the knowledge of the electron temperature \te, as done recently by our group (\citealt{Kennicutt:2003, Garnett:2004a,Bresolin:2004,Bresolin:2005}). Although this approach is potentially prone to systematic errors due to the effects of temperature gradients within the nebulae at high abundance (\citealt{Stasinska:2005}), the data suggest that the direct abundances are reliable at least up to the solar value (\citealt{Bresolin:2007}).

A further issue in nebular astrophysics is represented by the so-called `abundance discrepancy', consisting in the fact that when {\sc o\,ii} recombination lines are used to determine \opp\/ abundances in place of the much stronger \oiii\/ collisionally excited lines, the resulting total oxygen abundances are larger, by 0.2 dex on average (\citealt{Garcia-Rojas:2007a}). The effect is measured in Galactic \hii\/ regions, ionized by one or few stars, such as the Orion nebula, M8 and M17 (\citealt{Esteban:2004,Garcia-Rojas:2007}), as well as in more luminous extragalactic \hii\/ regions, such as 30 Dor in the LMC (\citealt{Peimbert:2003}), NGC~604 in M33 and others (\citealt{Esteban:2002,Esteban:2009}). As shown by \citet{Garcia-Rojas:2007a}, the effect does not seem to depend on metallicity or other nebular parameters, and can be interpreted with the presence of temperature fluctuations in \hii\/ regions (\citealt{Peimbert:1967}). The size of the discrepancy is large enough that it would affect the comparison 
between nebular and stellar abundances in such a way that it should be possible to discern which nebular abundance determination method best agrees with the stellar measurements.\\

In this paper we reassess the issue of present-day chemical abundances determined jointly from \hii\/ regions and massive stars by analyzing new \hii\/ region spectroscopy obtained in the southern spiral galaxy NGC~300.
This Scd galaxy is located in the Sculptor Group at a Cepheid distance of 1.88~Mpc (\citealt{Gieren:2005}).
Table~\ref{parameters} summarizes some essential parameters of this galaxy.
Its proximity has made it the subject of a number of investigations during the past decade concerning 
the stellar populations in the central/nuclear regions  (\citealt{Davidge:1998,Walcher:2006})
and in the outskirts  (\citealt{Tikhonov:2005,Bland-Hawthorn:2005,Mouhcine:2006,Vlajic:2009}), 
the star formation history (\citealt{Butler:2004}),
Wolf-Rayet stars and OB associations (\citealt{Schild:2003,Pietrzynski:2001}), 
planetary nebulae (\citealt{Soffner:1996,Pena:2009}),
variable stars (\citealt{Pietrzynski:2002,Mennickent:2004}),
dust content (\citealt{Helou:2004,Roussel:2005}),
X-ray sources (\citealt{Read:2001,Carpano:2005}),
supernova remnants (\citealt{Blair:1997,Pannuti:2000,Payne:2004})
and UV emission properties (\citealt{Munoz-Mateos:2007}).

Spectroscopic work on the blue supergiants of NGC~300 has recently been carried out by our group. Following the spectral classification of nearly 70 stars by \citet{Bresolin:2002a}, \citet{Urbaneja:2005} measured O, Mg and Si abundances for six early-B supergiants, while \citet{Kudritzki:2008} derived metallicities for 24 B8-A4 supergiants. \hii\/ region abundances have been first measured by \citet{Pagel:1979} in their seminal paper that introduced the \rtwothree\/ abundance indicator. Subsequent abundance studies were carried out by 
\citet{Webster:1983}, \citet{Edmunds:1984}, \citet{Deharveng:1988} and \citet{Christensen:1997}.
Additional nebular spectroscopic work includes various searches for Wolf-Rayet (W-R) stars (\citealt{Dodorico:1983, Schild:1992, Schild:2003}, among others), and the search for supernova remnants by \citet{Blair:1997}. \citet{Deharveng:1988} presented a catalog of 176 \hii\/ regions in NGC~300,
superseding a previous one by \citet{Sersic:1966}, and discussed the radial abundance gradient from a
compilation of new and previously published spectral data for 28 \hii\/ regions.


\begin{figure*}
\medskip
\center \includegraphics[width=1.\textwidth]{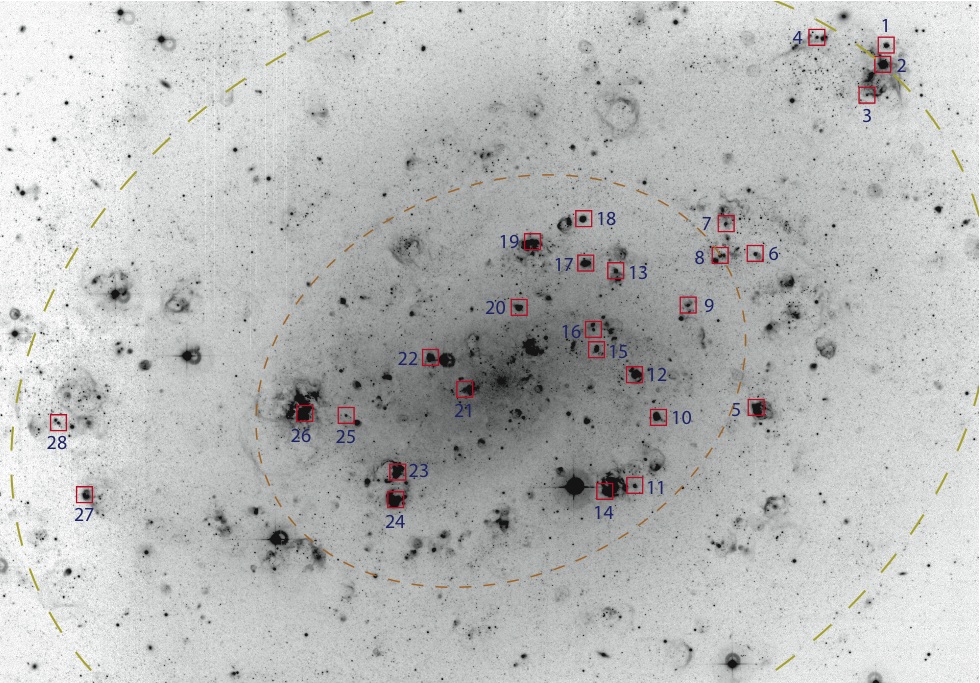}\medskip
\caption{Location of the \hii\/ regions studied in this work on a narrow-band \halpha\/ image of NGC~300 (courtesy ESO).
The \hii\/ regions are numbered in order of increasing right ascension. The dashed lines represent the location of the 
projected 0.5\,\rtf\/ and \rtf\/ radii (\rtf\,=\,9\farcm75).\\ \\ \label{image}}
\epsscale{1}
\end{figure*}

Despite all of these studies, until now only two \hii\/ regions in NGC~300 had a direct, \oiii\lin4363-based determination of the oxygen abundance (one each in \citealt{Pagel:1979} and \citealt{Webster:1983}),
limiting considerably the accuracy with which the metallicity gradient is known for this galaxy.
The availability of metallicity determinations for a large number of stars in NGC~300 (\citealt{Urbaneja:2005,Kudritzki:2008}) provided the motivation to carry out a modern re-evaluation of the nebular abundances in this galaxy, with the goal of measuring high-quality, direct abundances from the detection of the auroral lines. The data presented here allow a new comparison between stellar and \te-based \hii\/ region abundances, only the third of its kind for a spiral galaxy, after the Milky Way and M33. As we will show, this kind of comparisons has important implications, for example concerning the calibration of strong-line abundance methods. 

Our paper is organized as follows: the new observations and the data reduction are presented in \S~2, followed 
by the determination of electron temperatures of the ionized gas from the auroral lines in \S~3. Chemical abundances are derived in \S~4, and we discuss the abundance gradient in NGC~300, comparing \hii\/ regions with young stars, in \S~5. The discussion in \S~6 focuses on results of similar comparisons in other galaxies, and presents comments on the strong-line methods, the W-R star content, and the properties of the ionizing radiation in a wider sample of \hii\/ regions. We summarize our main results in \S~7.

\begin{deluxetable}{lc}
\tabletypesize{\footnotesize}
\tablecolumns{2}
\tablecaption{NGC~300: Galaxy parameters\label{parameters}}

\tablehead{
\colhead{\phantom{}Parameter\phantom{aaaaaaaaaaaaaaaaaaaaaaaaaaaaaaa}}	     &
\colhead{\phantom{}Value\phantom{}}       }
\startdata
\\[-2mm]
R.A. (J2000.0)\dotfill		& 	00:54:53.48 				\\
Decl. (J2000.0)\dotfill		&	$-$37:41:03.8			\\
Morphological type\dotfill					&	Scd					\\
Distance\dotfill				&	1.88 Mpc\tablenotemark{a}					\\
\rtf	\dotfill					&	9\farcm75 (5.33~kpc)				\\		
Inclination\dotfill			&	39\fdg8					\\					
Position angle of major axis\dotfill		&	114\fdg3					\\		
Heliocentric radial velocity\dotfill				&	144 km\,s$^{-1}$						\\
B$_T^0$\dotfill						&	8.49	\tablenotemark{b}							\\
M$_B^0$\dotfill						&	$-$17.88		
\enddata 
\tablecomments{All parameters from the HyperLeda database (\citealt{Paturel:2003}), except where noted. M$_B^0$ calculated from B$_T^0$ and the adopted distance.}
\tablenotetext{a}{\citet{Gieren:2005}}
\tablenotetext{b}{\citet{de-Vaucouleurs:1991}\\}
\end{deluxetable}

\section{Observations and data reduction}

Multi-object spectroscopy of \hii\/ regions in NGC~300 was obtained on the nights of August 17 and 18, 2006 at the 
Very Large Telescope of the European Southern Observatory on Cerro Paranal. The FORS2 instrument was used with 1~arcsec slits to cover targets in three different $6\farcm8\times6\farcm8$ regions of the galaxy, located near the galactic center (with two separate multi-object setups), as well as east and north-west of the center (one setup each). 
The targets had been selected from 
narrow-band \halpha\/ images obtained on July 1, 2006 with the same instrument.
The sky conditions were clear, and the seeing varied 
between 0.5 and 0.9 arcsec during most of the observing run (the seeing was 1.3 arcsec at the end of the run, while observing additional \hii\/ regions in the central field). 

Because of the requirement of covering a fairly  extended wavelength range (from \oii\lin3727 to \siii\llin9069,9532) at a moderate spectral resolution (5-10~\AA), we combined each object mask with three different grisms: 
600B ($4\times1800$s exposures, approximate range: 3500-6000~\AA\/ for a slit located near the center of the detector, 4.5~\AA\/ FWHM spectral resolution);
600R ($4\times1800$s, approximate range: 5300-8300~\AA, 5~\AA\/ resolution) and
300I ($2\times900$s, approximate range: 6100-10000~\AA, 10~\AA\/ resolution). 

The first stages of the data reduction, including bias subraction, flat field correction and wavelength calibration, were carried out with the EsoRex pipeline provided by the European Southern Observatory. Cosmic rays were removed 
with the {\sc L.A.Cosmic} routine (\citealt{van-Dokkum:2001}), while standard {\sc iraf}\footnote{{\sc iraf} is distributed by the National Optical Astronomy
Observatories, which are operated by the Association of Universities
for Research in Astronomy, Inc., under cooperative agreement with the
National Science Foundation.} tasks were used for the spectral extractions, image coadditions and flux calibration
(we obtained spectra of the  spectrophotometric standards BPM\,16274, EG\,21, EG\,274, LDS\,749B and NGC\,7293).
 
\begin{deluxetable*}{ccccl}
\tabletypesize{\scriptsize}
\tabletypesize{\footnotesize}
\tablecolumns{5}
\tablewidth{0pt}
\tablecaption{H\,\scriptsize II \small region sample\label{sample}}

\tablehead{
\colhead{\phantom{aaaaa}ID\phantom{aaaaa}}	     &
\colhead{\phantom{aaaaa}R.A.\phantom{aaaaa}}	 &
\colhead{\phantom{aaaaa}Decl.\phantom{aaaaa}}  &
\colhead{\phantom{aaa}R/R$_{25}$\phantom{aaa}}	 &
\colhead{\phantom{aa}ID\phantom{aa}}	 \\[0.5mm]
\colhead{}       &
\colhead{(J2000.0)}   &
\colhead{(J2000.0)}   &
\colhead{}   &
\colhead{} \\[1mm]
\colhead{(1)}	&
\colhead{(2)}	&
\colhead{(3)}	&
\colhead{(4)}	&
\colhead{(5)}	}
\startdata
\\[-2mm]
1\dotfill  & 00~ 54~ 16.22  &  $-$37~ 34~ 35.9  &   1.03 &	5 \hfill(W\,16)      \\       
2\dotfill  & 00~ 54~ 16.28  &  $-$37~ 34~ 54.9  &   1.01 &	6 \hfill(P\,7; W\,14-15; D\,1)      \\       
3\dotfill  & 00~ 54~ 17.98  &  $-$37~ 35~ 33.2  &   0.93 &	\nodata      \\       
4\dotfill  & 00~ 54~ 22.83  &  $-$37~ 34~ 25.9  &   0.97 &	12     \\       
5\dotfill  & 00~ 54~ 28.70  &  $-$37~ 41~ 32.8  &   0.55 &	24 \hfill(W\,11; D\,2; S\,13)   \\       
6\dotfill  & 00~ 54~ 28.83  &  $-$37~ 38~ 36.3  &   0.56 &	25      \\       
7\dotfill  & 00~ 54~ 31.70  &  $-$37~ 38~ 01.5  &   0.55 &	30      \\       
8\dotfill  & 00~ 54~ 32.20  &  $-$37~ 38~ 37.7  &   0.50 &	32 \hfill(S\,19)      \\       
9\dotfill  & 00~ 54~ 35.35  &  $-$37~ 39~ 35.3  &   0.40 &	37      \\       
10\dotfill  & 00~ 54~ 38.16  &  $-$37~ 41~ 44.8  &   0.36 &	39 \hfill(S\,21)     \\       
11\dotfill  & 00~ 54~ 40.53  &  $-$37~ 43~ 03.1  &   0.41 &	46 \hfill(S\,23)     \\       
12\dotfill  & 00~ 54~ 40.56  &  $-$37~ 40~ 55.3  &   0.27 &	45 \hfill(W\,4; D\,3)      \\       
13\dotfill  & 00~ 54~ 42.29  &  $-$37~ 38~ 56.1  &   0.33 &	50      \\       
14\dotfill  & 00~ 54~ 43.42  &  $-$37~ 43~ 09.3  &   0.38 &	53A\phantom{aaa} \hfill(P\,5; W\,5; D\,6; S\,29)     \\       
15\dotfill  & 00~ 54~ 44.22  &  $-$37~ 40~ 25.7  &   0.20 &	56      \\       
16\dotfill  & 00~ 54~ 44.52  &  $-$37~ 40~ 03.1  &   0.21 &	58      \\       
17\dotfill  & 00~ 54~ 45.28  &  $-$37~ 38~ 46.5  &   0.31 &	61 \hfill(S\,31)     \\       
18\dotfill  & 00~ 54~ 45.37  &  $-$37~ 37~ 56.1  &   0.41 &	63      \\       
19\dotfill  & 00~ 54~ 50.34  &  $-$37~ 38~ 22.6  &   0.34 &	77 \hfill(P\,1; D\,7; S\,38)     \\       
20\dotfill  & 00~ 54~ 51.69  &  $-$37~ 39~ 38.7  &   0.18 &	84 \hfill(S\,41)     \\       
21\dotfill  & 00~ 54~ 57.02  &  $-$37~ 41~ 12.4  &   0.08 &	\nodata      \\       
22\dotfill  & 00~ 55~ 00.38  &  $-$37~ 40~ 35.4  &   0.17 &	109 \hfill(P\,2; D\,9; S\,53)     \\       
23\dotfill  & 00~ 55~ 03.56  &  $-$37~ 42~ 48.5  &   0.28 &	118A \hfill(D\,11; S\,59)     \\       
24\dotfill  & 00~ 55~ 03.65  &  $-$37~ 43~ 19.4  &   0.33 &	119A \hfill(W\,6; D\,12; S\,58)     \\       
25\dotfill  & 00~ 55~ 08.53  &  $-$37~ 41~ 42.3  &   0.32 &	128      \\       
26\dotfill  & 00~ 55~ 12.58  &  $-$37~ 41~ 39.2  &   0.40 &	137A \hfill(W\,7; D\,14)      \\       
27\dotfill  & 00~ 55~ 33.94  &  $-$37~ 43~ 12.8  &   0.86 &	159 \hfill(P\,4; S\,80)     \\       
28\dotfill  & 00~ 55~ 36.45  &  $-$37~ 41~ 49.7  &   0.91 &	161             
\enddata
\tablecomments{Units of right ascension are hours, minutes and seconds, and units of declination
are degrees, arcminutes and arcseconds. Col.~(1): \hii\/ region identification. Col.~(2): Right
Ascension. Col.~(3): Declination. Col.~(4): Galactocentric distance in units of R$_{25}$\,=\,9.75 arcmin.
Col.~(5): Identification from \citet{Deharveng:1988}. Additional IDs in brackets - P: \citet{Pagel:1979}, \citet{Edmunds:1984};
W: \citet{Webster:1983}; D: \citet{Dodorico:1983}; S: \citet{Sersic:1966}\\}
\end{deluxetable*}

Since in this paper we focus on direct measurements of the \hii\/ region chemical abundances, without relying on statistical abundance indicators, we limit our discussion to the 28 targets for which we could measure reliable electron temperatures from auroral-to-nebular line ratios, with the methods explained in \S~3. The spatial distribution of this sample is shown in Fig.~\ref{image}, 
while the celestial coordinates of the targets and their identification from previous studies are summarized in 
Table~\ref{sample}. The positions were measured on an \halpha\/ image taken by FB in August 2000  at the MPG/ESO 2.2m telescope on La Silla, equipped with the Wide Field Imager. The rms uncertainty of the astrometric solution, derived using 
stars in the USNO catalog, is 0.4 arcsec.

The deprojected galactocentric distances in Col.~4 are given in terms of the 25th magnitude $B$-band isophotal radius, \rtf\,=\,9\farcm75 (=\,5.33~kpc at the adopted distance of 1.88~Mpc), and were calculated adopting 
the disk parameters reported in Table~\ref{parameters}.
As Table~\ref{sample} shows, almost all of our targets are included in the catalog compiled by \citet[entries from this catalog will be indicated with the prefix De]{Deharveng:1988}. 
The only exceptions are our objects \#3 and \#21, both rather compact \hii\/ regions (\#21 could be the eastern extension of what \citealt{Deharveng:1988} considered as a single object, De\,100 in their catalog).

The reddening-corrected emission lines presented in Tables~\ref{fluxes_a} and \ref{fluxes_b} were measured with the {\tt splot} task in {\sc iraf}. The reddening correction was obtained with an iterative procedure from the Balmer decrement, assuming case B \hi\/ line ratios (\citealt{Hummer:1987}), calculated at the electron temperatures derived from the auroral lines, simultaneously determining the correction for the underlying stellar  absorption. 
The interstellar reddening law of \citet{Seaton:1979} has been adopted.
The internal consistency of the flux calibration in the blue was checked by ensuring that the strengths of the higher order lines of the Balmer series that are still measurable in our spectra (H9-H11) agree, within the errors, with the case B predictions.

The line flux errors were calculated using the expression by \citet{Gonzalez-Delgado:1994}:
\begin{equation}
\sigma_l = \sigma_{\rm cont}\, N^{1/2}\, [1 + EW/(N\Delta)]^{1/2}
\end{equation}
 
\noindent
where $\sigma_{\rm cont}$ is the standard deviation of the continuum near the emission line, $N$ is the width of the region used to measure the line in pixels, $\Delta$ is the spectral dispersion in \AA\,pixel$^{-1}$, and EW represents the equivalent width of the line. The final errors quoted in Tables~\ref{fluxes_a} and \ref{fluxes_b} include, added in quadrature, contributions of 1\% and 
4\% due to the uncertainties in the flat fielding and the flux calibration, respectively, as well as the uncertainty in 
scaling spectra obtained through different grisms (estimated to be 3\%). In propagating the errors we also accounted for the uncertainty in the extinction coefficient c(H$\beta$), which amounts to about 0.07 mag.
The \hbeta\/ line flux that appears in Col.~9 of Table~\ref{fluxes_b} has been corrected for extinction, and should be regarded as a lower limit of the real flux, due to slit losses.
The missing entries for the \hei\lin5876 and/or \siii\lin6312 lines are due to the fact that the 600R grism introduces a shift along the spatial direction, so that spectra near the top edge of the two 2048$\times$2048 FORS2 detectors cannot be recorded.
In one case \nii\lin5755 is also missing, because the line could not be measured in the blue (600B) spectra, while in all of the remaining targets this line was measured in either the red (600R) or the blue spectra, or in both.
 
Only one of our targets, \#27 (=\,De\,159), appears in the list of optically-selected supernova remnants (SNRs)
by \citet{Blair:1997} as their source NGC300-S28. We detect a modest \oi\llin6300,6360 emission (\oi/H$\alpha$\,=\,0.04), but the shock excitation diagnostic line ratios \oi/H$\alpha$ and \sii/H$\alpha$ that we measure are 3 and 2.5 times
smaller than the \citet{Blair:1997}'s values, respectively. 
We find \sii/H$\alpha$\,=\,0.29, considerably smaller than the minimum value of 0.4 commonly adopted in SNR searches
to discriminate against photoionized nebulae, but this is also the largest \sii/H$\alpha$ ratio in our sample. 
This \hii\/ region appears as being composed of two separate bright knots, so it is likely that the object centered in our slit does not correspond to the one observed by \citet{Blair:1997}. In fact, while we do detect a few 
faint emission lines, such as \mgi\lin4562 and \feiii\lin4658 and \lin5270, found in the spectra of SNRs (e.g.~\citealt{Osterbrock:1973}), this spectrum would not be flagged as that of a SNR on the basis of a \sii/H$\alpha$ \vs\/ \nii/H$\alpha$ diagnostic diagram (\citealt{Sabbadin:1977}). We conclude that the SNR spectrum is fairly diluted by the \hii\/ region spectrum, and we keep this target in our analysis.
A number of additional SNRs within \hii\/ regions have been discovered in NGC~300 with radio continuum observations by \citet{Pannuti:2000} and \citet{Payne:2004}. Among these are several \hii\/ regions in our sample (D39, D53A, D61, D77, D84, D109, D118A, D119A, D137A and D159). From our line ratios we do not find evidence for significant shock excitation in these targets.\vspace{0.3cm}

\begin{deluxetable*}{cccccccccc}
\tabletypesize{\scriptsize}
\tablecolumns{10}
\tablewidth{0pt}
\tablecaption{Reddening-corrected fluxes (A)\label{fluxes_a}}

\tablehead{
\colhead{\phantom{aaaaa}ID\phantom{aaaaa}}	     &
\colhead{\oii}	 &
\colhead{\neiii}	 &
\colhead{\sii}	 &
\colhead{\oiii}	 &
\colhead{\oiii}	 &
\colhead{\nii}	 &
\colhead{\hei}	 &
\colhead{\siii} &
\colhead{\nii}   \\[0.5mm]
\colhead{}       &
\colhead{3727}   &
\colhead{3868}   &
\colhead{4072}   &
\colhead{4363}   &
\colhead{5007}   &
\colhead{5755} &
\colhead{5876}  &
\colhead{6312}     &
\colhead{6583}     \\[1mm]
\colhead{(1)}	&
\colhead{(2)}	&
\colhead{(3)}	&
\colhead{(4)}	&
\colhead{(5)}	&
\colhead{(6)}	&
\colhead{(7)}   &
\colhead{(8)}   &
\colhead{(9)}   &
\colhead{(10)}    }
\startdata
\\[-2mm]
 1\dotfill &   243 $\pm$   17 &    20.6 $\pm$  1.4 &    2.31 $\pm$ 0.23 &    3.38 $\pm$ 0.24 &     321 $\pm$   19 &    \nodata        &     \nodata        &     \nodata        &    13.5 $\pm$  0.9 \\  
 2\dotfill &   166 $\pm$   12 &    33.8 $\pm$  2.3 &    1.39 $\pm$ 0.11 &    5.59 $\pm$ 0.35 &     504 $\pm$   29 &   0.18 $\pm$ 0.03 &    12.5 $\pm$  0.8 &    1.85 $\pm$ 0.13 &     7.7 $\pm$  0.5 \\  
 3\dotfill &   373 $\pm$   27 &     4.6 $\pm$  0.5 &    3.46 $\pm$ 0.37 &    1.16 $\pm$ 0.19 &     119 $\pm$    7 &   0.00 $\pm$ 0.15 &    11.5 $\pm$  0.8 &    0.00 $\pm$ 0.11 &    23.4 $\pm$  1.6 \\  
 4\dotfill &   286 $\pm$   20 &    14.9 $\pm$  1.0 &    2.20 $\pm$ 0.19 &    2.14 $\pm$ 0.17 &     253 $\pm$   15 &   0.00 $\pm$ 0.12 &    13.0 $\pm$  0.8 &    1.73 $\pm$ 0.16 &    20.4 $\pm$  1.3 \\  
 5\dotfill &   296 $\pm$   21 &    10.2 $\pm$  1.1 &    3.25 $\pm$ 0.42 &    1.10 $\pm$ 0.15 &     226 $\pm$   13 &   0.00 $\pm$ 0.08 &    11.9 $\pm$  0.8 &    1.34 $\pm$ 0.16 &    29.7 $\pm$  2.0 \\  
 6\dotfill &   275 $\pm$   20 &    13.9 $\pm$  1.0 &    3.23 $\pm$ 0.25 &    1.16 $\pm$ 0.14 &     246 $\pm$   14 &   0.00 $\pm$ 0.12 &    13.2 $\pm$  0.9 &    1.38 $\pm$ 0.16 &    27.2 $\pm$  1.8 \\  
 7\dotfill &   134 $\pm$   10 &    75.1 $\pm$  5.3 &    3.64 $\pm$ 0.27 &    9.07 $\pm$ 0.59 &     908 $\pm$   53 &   0.00 $\pm$ 0.08 &     \nodata        &     \nodata        &    15.6 $\pm$  1.1 \\  
 8\dotfill &   307 $\pm$   22 &     7.2 $\pm$  0.6 &    2.23 $\pm$ 0.17 &    0.77 $\pm$ 0.10 &     144 $\pm$    8 &   0.50 $\pm$ 0.08 &    12.1 $\pm$  0.8 &    1.32 $\pm$ 0.12 &    34.1 $\pm$  2.2 \\  
 9\dotfill &   172 $\pm$   12 &    12.0 $\pm$  0.8 &    1.95 $\pm$ 0.15 &    0.89 $\pm$ 0.08 &     235 $\pm$   14 &   0.00 $\pm$ 0.04 &    13.1 $\pm$  0.8 &    1.09 $\pm$ 0.08 &    32.5 $\pm$  2.1 \\  
10\dotfill &   180 $\pm$   13 &    13.1 $\pm$  0.9 &    1.13 $\pm$ 0.09 &    0.91 $\pm$ 0.07 &     236 $\pm$   14 &   0.23 $\pm$ 0.02 &    11.0 $\pm$  0.7 &    1.11 $\pm$ 0.08 &    19.2 $\pm$  1.3 \\  
11\dotfill &   258 $\pm$   19 &    10.0 $\pm$  0.7 &    2.74 $\pm$ 0.26 &    1.11 $\pm$ 0.18 &     201 $\pm$   12 &   0.00 $\pm$ 0.07 &    11.5 $\pm$  0.8 &    1.34 $\pm$ 0.12 &    29.3 $\pm$  1.9 \\  
12\dotfill &   262 $\pm$   19 &     0.0 $\pm$  0.2 &    2.80 $\pm$ 0.25 &    0.00 $\pm$ 0.11 &      30 $\pm$    2 &   0.37 $\pm$ 0.09 &     8.6 $\pm$  0.6 &    0.73 $\pm$ 0.07 &    54.1 $\pm$  3.5 \\  
13\dotfill &   246 $\pm$   18 &     3.9 $\pm$  0.3 &    2.06 $\pm$ 0.38 &    0.00 $\pm$ 0.21 &     119 $\pm$    7 &   0.63 $\pm$ 0.08 &    13.1 $\pm$  0.9 &    1.04 $\pm$ 0.09 &    40.2 $\pm$  2.6 \\  
14\dotfill &   248 $\pm$   18 &     8.1 $\pm$  0.6 &    1.78 $\pm$ 0.12 &    0.75 $\pm$ 0.06 &     181 $\pm$   11 &   0.40 $\pm$ 0.03 &    11.6 $\pm$  0.8 &    1.40 $\pm$ 0.10 &    41.1 $\pm$  2.7 \\  
15\dotfill &   185 $\pm$   13 &     2.6 $\pm$  0.2 &    2.22 $\pm$ 0.19 &    0.00 $\pm$ 0.11 &      92 $\pm$    5 &   0.57 $\pm$ 0.11 &    12.5 $\pm$  0.8 &    1.10 $\pm$ 0.10 &    58.1 $\pm$  3.8 \\  
16\dotfill &   264 $\pm$   19 &     3.4 $\pm$  0.3 &    3.93 $\pm$ 0.30 &    0.00 $\pm$ 0.34 &      95 $\pm$    6 &   0.69 $\pm$ 0.14 &    11.2 $\pm$  0.8 &    0.99 $\pm$ 0.12 &    54.1 $\pm$  3.6 \\  
17\dotfill &   213 $\pm$   15 &     6.6 $\pm$  0.5 &    1.29 $\pm$ 0.09 &    0.59 $\pm$ 0.05 &     192 $\pm$   11 &   0.34 $\pm$ 0.04 &    13.6 $\pm$  0.9 &    1.32 $\pm$ 0.11 &    27.6 $\pm$  1.8 \\  
18\dotfill &   259 $\pm$   18 &     4.6 $\pm$  0.3 &    2.10 $\pm$ 0.17 &    0.00 $\pm$ 0.10 &     118 $\pm$    7 &   0.41 $\pm$ 0.06 &    11.1 $\pm$  0.7 &    1.26 $\pm$ 0.10 &    46.8 $\pm$  3.1 \\  
19\dotfill &   192 $\pm$   14 &     6.8 $\pm$  0.5 &    2.17 $\pm$ 0.26 &    0.61 $\pm$ 0.05 &     165 $\pm$   10 &   0.25 $\pm$ 0.03 &    15.0 $\pm$  1.0 &    0.95 $\pm$ 0.07 &    23.3 $\pm$  1.5 \\  
20\dotfill &   146 $\pm$   10 &     9.9 $\pm$  0.7 &    1.15 $\pm$ 0.09 &    0.71 $\pm$ 0.06 &     227 $\pm$   13 &   0.28 $\pm$ 0.03 &    12.8 $\pm$  0.8 &    1.15 $\pm$ 0.08 &    23.9 $\pm$  1.6 \\  
21\dotfill &   217 $\pm$   16 &     3.3 $\pm$  0.4 &    3.42 $\pm$ 0.31 &    0.00 $\pm$ 0.09 &     103 $\pm$    6 &   0.63 $\pm$ 0.08 &     9.6 $\pm$  0.7 &    1.09 $\pm$ 0.12 &    72.7 $\pm$  4.8 \\  
22\dotfill &   270 $\pm$   19 &     3.3 $\pm$  0.3 &    3.42 $\pm$ 0.34 &    0.00 $\pm$ 0.16 &      77 $\pm$    4 &   0.47 $\pm$ 0.06 &    10.6 $\pm$  0.7 &    0.82 $\pm$ 0.08 &    61.5 $\pm$  4.0 \\  
23\dotfill &   176 $\pm$   13 &     9.8 $\pm$  0.7 &    1.61 $\pm$ 0.13 &    0.57 $\pm$ 0.07 &     198 $\pm$   12 &   0.28 $\pm$ 0.02 &    11.7 $\pm$  0.8 &    0.99 $\pm$ 0.09 &    27.8 $\pm$  1.8 \\  
24\dotfill &   197 $\pm$   14 &     5.2 $\pm$  0.4 &    1.12 $\pm$ 0.11 &    0.62 $\pm$ 0.05 &     180 $\pm$   10 &   0.37 $\pm$ 0.03 &    12.6 $\pm$  0.8 &     \nodata        &    32.4 $\pm$  2.1 \\  
25\dotfill &   287 $\pm$   21 &     4.2 $\pm$  0.3 &    2.15 $\pm$ 0.20 &    0.00 $\pm$ 0.26 &     133 $\pm$    8 &   0.60 $\pm$ 0.12 &    10.4 $\pm$  0.7 &    1.07 $\pm$ 0.13 &    44.0 $\pm$  2.9 \\  
26\dotfill &   160 $\pm$   11 &    11.2 $\pm$  0.8 &    1.77 $\pm$ 0.16 &    1.19 $\pm$ 0.10 &     259 $\pm$   15 &   0.27 $\pm$ 0.03 &    12.2 $\pm$  0.8 &    1.43 $\pm$ 0.10 &    20.3 $\pm$  1.3 \\  
27\dotfill &   357 $\pm$   25 &     9.8 $\pm$  0.7 &    5.44 $\pm$ 0.36 &    1.59 $\pm$ 0.11 &     178 $\pm$   10 &   0.54 $\pm$ 0.07 &    12.2 $\pm$  0.8 &    1.50 $\pm$ 0.11 &    31.8 $\pm$  2.1 \\  
28\dotfill &   314 $\pm$   22 &    12.7 $\pm$  0.9 &    2.04 $\pm$ 0.15 &    1.89 $\pm$ 0.13 &     244 $\pm$   14 &   0.43 $\pm$ 0.10 &    11.8 $\pm$  0.8 &    1.51 $\pm$ 0.19 &    21.3 $\pm$  1.4 \\  
\\[-6mm]
\enddata
\tablecomments{The line fluxes  are in units of H$\beta$\,=\,100.\\}
\end{deluxetable*}

\begin{deluxetable*}{cccccccccc}
\tabletypesize{\scriptsize}
\tablecolumns{10}
\tablewidth{0pt}
\tablecaption{Reddening-corrected fluxes (B)\label{fluxes_b}}

\tablehead{
\colhead{\phantom{aaaaa}ID\phantom{aaaaa}}	     &
\colhead{\hei}	 &
\colhead{\sii}	 &
\colhead{\ariii}	 &
\colhead{\oii}	 &
\colhead{\hi}	 &
\colhead{\siii}	 &
\colhead{\hi}	 &
\colhead{F(H$\beta$)} &
\colhead{$c$(\hbeta)}   \\[0.5mm]
\colhead{}       &
\colhead{6678}   &
\colhead{6717+6731}   &
\colhead{7135}   &
\colhead{7325}   &
\colhead{9015}   &
\colhead{9069} &
\colhead{9229}  &
\colhead{(erg\,s$^{-1}$\,cm$^{-2}$)}     &
\colhead{(mag)}     \\[1mm]
\colhead{(1)}	&
\colhead{(2)}	&
\colhead{(3)}	&
\colhead{(4)}	&
\colhead{(5)}	&
\colhead{(6)}	&
\colhead{(7)}   &
\colhead{(8)}   &
\colhead{(9)}   &
\colhead{(10)}    }
\startdata
\\[-2mm]
 1\dotfill &   2.9 $\pm$  0.2 &   22.1 $\pm$  1.0 &    7.3 $\pm$  0.5    &    6.1 $\pm$  0.4    &    1.9 $\pm$  0.2    &   17.0 $\pm$  1.1    &    2.7 $\pm$  0.2    &    7.6 $\times 10^{-15}$ &  0.23 \\   
 2\dotfill &   3.1 $\pm$  0.2 &   17.4 $\pm$  0.8 &    8.7 $\pm$  0.6    &    4.3 $\pm$  0.3    &    1.9 $\pm$  0.1    &   19.6 $\pm$  1.3    &    2.5 $\pm$  0.2    &    8.7 $\times 10^{-15}$ &  0.12 \\   
 3\dotfill &   2.8 $\pm$  0.2 &   52.0 $\pm$  2.5 &    5.3 $\pm$  0.4    &    8.0 $\pm$  0.6    &    2.3 $\pm$  0.2    &   12.7 $\pm$  0.9    &    2.5 $\pm$  0.2    &    1.0 $\times 10^{-15}$ &  0.03 \\   
 4\dotfill &   3.9 $\pm$  0.3 &   33.2 $\pm$  1.6 &    9.2 $\pm$  0.6    &    6.0 $\pm$  0.4    &    1.7 $\pm$  0.1    &   20.7 $\pm$  1.4    &    2.5 $\pm$  0.2    &    6.5 $\times 10^{-15}$ &  0.40 \\   
 5\dotfill &   3.3 $\pm$  0.2 &   52.7 $\pm$  2.5 &    8.8 $\pm$  0.6    &    4.7 $\pm$  0.3    &    2.0 $\pm$  0.1    &   26.4 $\pm$  1.7    &    2.5 $\pm$  0.2    &    6.9 $\times 10^{-15}$ &  0.15 \\   
 6\dotfill &   3.3 $\pm$  0.2 &   38.6 $\pm$  1.8 &    8.7 $\pm$  0.6    &    4.1 $\pm$  0.4    &    2.0 $\pm$  0.2    &   20.3 $\pm$  1.4    &    2.5 $\pm$  0.2    &    4.7 $\times 10^{-15}$ &  0.25 \\   
 7\dotfill &   3.1 $\pm$  0.3 &   34.9 $\pm$  1.7 &   14.7 $\pm$  1.0    &    3.8 $\pm$  0.3    &    1.7 $\pm$  0.2    &   30.9 $\pm$  2.1    &    2.7 $\pm$  0.2    &    2.7 $\times 10^{-15}$ &  0.00 \\   
 8\dotfill &   3.1 $\pm$  0.2 &   44.2 $\pm$  2.1 &    7.9 $\pm$  0.5    &    4.4 $\pm$  0.3    &    1.8 $\pm$  0.1    &   20.6 $\pm$  1.4    &    2.6 $\pm$  0.2    &    7.6 $\times 10^{-15}$ &  0.34 \\   
 9\dotfill &   3.5 $\pm$  0.2 &   33.1 $\pm$  1.5 &   10.3 $\pm$  0.7    &    3.2 $\pm$  0.2    &    1.9 $\pm$  0.1    &   27.2 $\pm$  1.8    &    2.6 $\pm$  0.2    &    7.2 $\times 10^{-15}$ &  0.33 \\   
10\dotfill &   3.0 $\pm$  0.2 &   16.3 $\pm$  0.8 &    9.5 $\pm$  0.6    &    2.2 $\pm$  0.2    &    1.7 $\pm$  0.1    &   26.5 $\pm$  1.7    &    2.6 $\pm$  0.2    &   42.0 $\times 10^{-15}$ &  0.72 \\   
11\dotfill &   3.5 $\pm$  0.2 &   41.8 $\pm$  2.0 &   10.7 $\pm$  0.7    &    5.6 $\pm$  0.4    &    2.2 $\pm$  0.2    &   23.8 $\pm$  1.6    &    2.5 $\pm$  0.2    &    6.6 $\times 10^{-15}$ &  0.20 \\   
12\dotfill &   2.5 $\pm$  0.2 &   72.5 $\pm$  3.4 &    4.6 $\pm$  0.3    &    2.8 $\pm$  0.2    &    2.1 $\pm$  0.1    &   22.0 $\pm$  1.4    &    2.6 $\pm$  0.2    &    9.1 $\times 10^{-15}$ &  0.17 \\   
13\dotfill &   2.9 $\pm$  0.2 &   37.5 $\pm$  1.8 &    6.7 $\pm$  0.4    &    2.8 $\pm$  0.2    &    1.8 $\pm$  0.1    &   21.2 $\pm$  1.4    &    2.6 $\pm$  0.2    &    9.0 $\times 10^{-15}$ &  0.43 \\   
14\dotfill &   3.8 $\pm$  0.2 &   41.6 $\pm$  1.9 &   12.1 $\pm$  0.8    &    4.7 $\pm$  0.3    &    2.0 $\pm$  0.1    &   29.6 $\pm$  1.9    &    2.6 $\pm$  0.2    &   48.5 $\times 10^{-15}$ &  0.17 \\   
15\dotfill &   3.5 $\pm$  0.2 &   48.0 $\pm$  2.3 &    9.0 $\pm$  0.6    &    3.0 $\pm$  0.2    &    1.9 $\pm$  0.1    &   28.1 $\pm$  1.8    &    2.6 $\pm$  0.2    &   17.1 $\times 10^{-15}$ &  0.43 \\   
16\dotfill &   3.1 $\pm$  0.2 &   53.0 $\pm$  2.5 &    8.3 $\pm$  0.6    &    3.8 $\pm$  0.3    &    2.1 $\pm$  0.2    &   24.3 $\pm$  1.6    &    2.6 $\pm$  0.2    &    6.9 $\times 10^{-15}$ &  0.50 \\   
17\dotfill &   3.3 $\pm$  0.2 &   27.3 $\pm$  1.3 &    8.8 $\pm$  0.6    &    2.3 $\pm$  0.2    &    1.9 $\pm$  0.1    &   27.3 $\pm$  1.8    &    2.6 $\pm$  0.2    &   22.5 $\times 10^{-15}$ &  0.28 \\   
18\dotfill &   3.6 $\pm$  0.2 &   57.2 $\pm$  2.7 &   10.6 $\pm$  0.7    &    5.2 $\pm$  0.4    &    1.6 $\pm$  0.1    &   25.9 $\pm$  1.7    &    2.6 $\pm$  0.3    &    9.0 $\times 10^{-15}$ &  0.13 \\   
19\dotfill &   3.8 $\pm$  0.2 &   30.6 $\pm$  1.4 &    6.3 $\pm$  0.4    &    2.1 $\pm$  0.1    &    2.0 $\pm$  0.1    &   23.1 $\pm$  1.5    &    2.6 $\pm$  0.2    &   31.2 $\times 10^{-15}$ &  0.24 \\   
20\dotfill &   3.6 $\pm$  0.2 &   20.0 $\pm$  0.9 &   10.8 $\pm$  0.7    &    2.3 $\pm$  0.2    &    1.9 $\pm$  0.1    &   28.6 $\pm$  1.9    &    2.5 $\pm$  0.2    &   44.1 $\times 10^{-15}$ &  0.33 \\   
21\dotfill &   3.0 $\pm$  0.2 &   55.5 $\pm$  2.6 &   10.2 $\pm$  0.7    &    3.5 $\pm$  0.3    &    1.8 $\pm$  0.2    &   28.8 $\pm$  1.9    &    2.6 $\pm$  0.2    &   11.1 $\times 10^{-15}$ &  0.76 \\   
22\dotfill &   3.0 $\pm$  0.2 &   71.6 $\pm$  3.3 &    7.6 $\pm$  0.5    &    3.7 $\pm$  0.3    &    1.9 $\pm$  0.1    &   22.3 $\pm$  1.5    &    2.5 $\pm$  0.2    &    8.3 $\times 10^{-15}$ &  0.29 \\   
23\dotfill &   3.4 $\pm$  0.2 &   28.0 $\pm$  1.3 &   10.5 $\pm$  0.7    &    3.0 $\pm$  0.2    &    2.0 $\pm$  0.1    &   24.0 $\pm$  1.6    &    2.5 $\pm$  0.2    &   45.0 $\times 10^{-15}$ &  0.31 \\   
24\dotfill &   2.9 $\pm$  0.2 &   30.3 $\pm$  1.4 &    9.3 $\pm$  0.6    &    3.6 $\pm$  0.2    &    2.1 $\pm$  0.1    &   26.5 $\pm$  1.7    &    2.6 $\pm$  0.2    &   27.5 $\times 10^{-15}$ &  0.04 \\   
25\dotfill &   3.4 $\pm$  0.2 &   45.0 $\pm$  2.1 &   10.7 $\pm$  0.7    &    5.7 $\pm$  0.4    &    2.3 $\pm$  0.2    &   24.6 $\pm$  1.6    &    2.6 $\pm$  0.2    &    8.4 $\times 10^{-15}$ &  0.47 \\   
26\dotfill &   3.7 $\pm$  0.2 &   22.1 $\pm$  1.0 &   11.4 $\pm$  0.7    &    3.2 $\pm$  0.2    &    1.7 $\pm$  0.1    &   28.2 $\pm$  1.8    &    2.5 $\pm$  0.2    &   30.5 $\times 10^{-15}$ &  0.33 \\   
27\dotfill &   2.7 $\pm$  0.2 &   81.7 $\pm$  3.8 &    6.4 $\pm$  0.4    &    6.0 $\pm$  0.4    &    1.3 $\pm$  0.1    &   17.3 $\pm$  1.1    &    2.6 $\pm$  0.2    &    7.7 $\times 10^{-15}$ &  0.22 \\   
28\dotfill &   4.2 $\pm$  0.2 &   37.2 $\pm$  1.7 &    7.8 $\pm$  0.6    &    6.0 $\pm$  0.4    &    1.7 $\pm$  0.1    &   21.2 $\pm$  1.4    &    2.7 $\pm$  0.2    &    4.0 $\times 10^{-15}$ &  0.16 \\   
\\[-6mm]
\enddata
\tablecomments{The line fluxes  are in units of H$\beta$\,=\,100.\\}
\end{deluxetable*}

\section{Electron Temperatures}
The knowledge of the electron temperature of \hii\/ regions is crucial to obtain reliable nebular chemical abundances,
because of the exponential temperature dependence of the line emissivity. In typical extragalactic work the most common auroral line used for \te\/ determinations, \oiii\lin4363, becomes very weak at moderately large oxygen abundance, as a result of the increasing cooling of the gas with metallicity. 
Somewhat surprisingly, the only previous detections of this line in NGC~300 are those in region 7 of \citet[our region 2\,=\,De\,6]{Pagel:1979} and
in region 7 of \citet[our region 26\,=\,De\,137A]{Webster:1983}.
Our much deeper spectra allowed us to detect \oiii\lin4363 in 20 \hii\/ regions. Moreover, it has been shown that 
for faint extragalactic \hii\/ regions the auroral lines of other ions, in particular \siii\lin6312 and \nii\lin5755, can be used as surrogates of \oiii\lin4363
when the latter is not detected, especially at gas metallicities approaching the solar one (\citealt{Bresolin:2004,Bresolin:2005}), albeit the derived electron temperatures and chemical abundances are somewhat model-dependent, and
carry larger uncertainties. We have detected the \siii\lin6312 auroral line in 8 additional objects, and we derived 
electron temperatures for the O$^{++}$ emitting region from this line, as explained below.


\begin{figure}
\medskip
\center \includegraphics[width=0.47\textwidth]{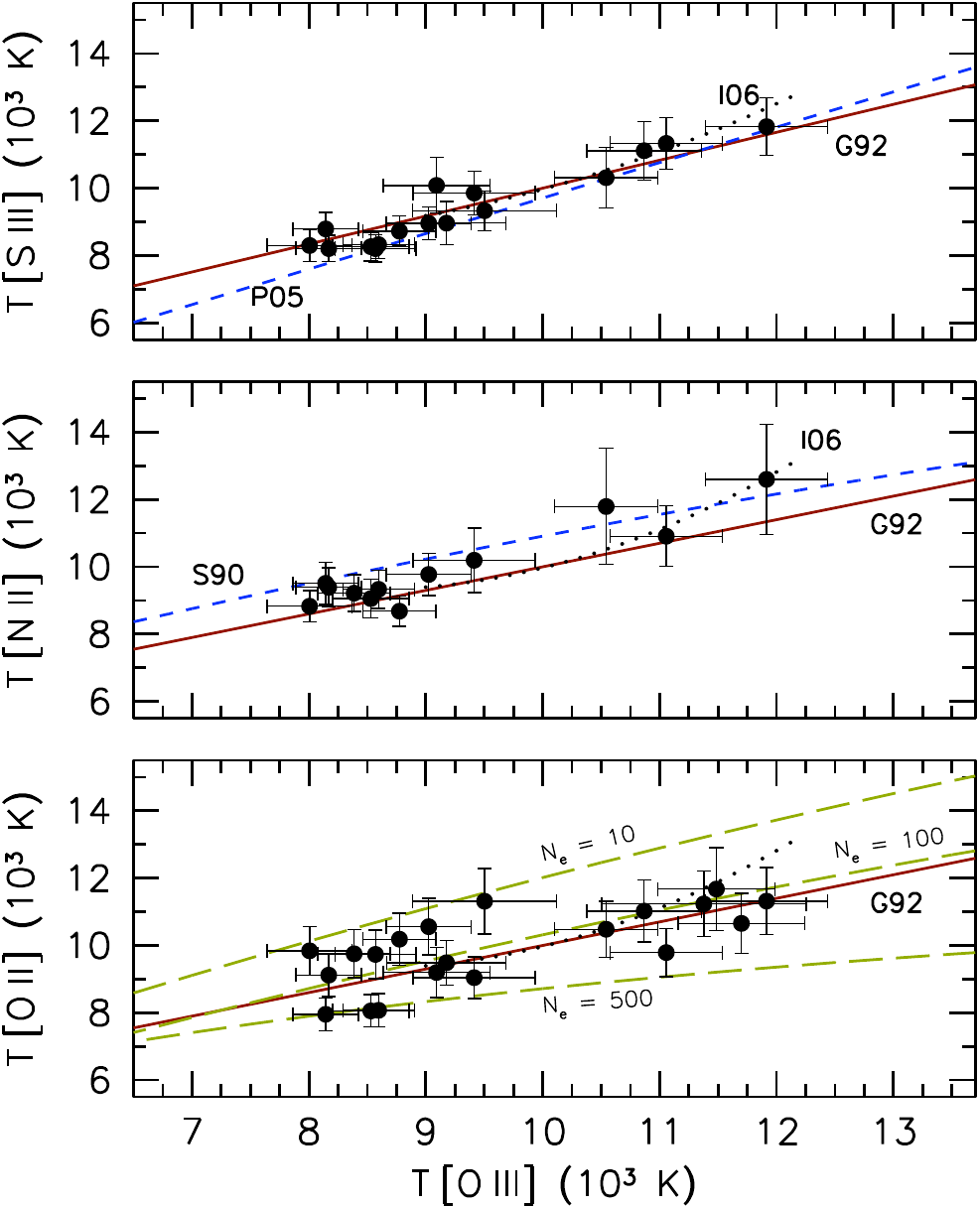}\medskip
\caption{Relations between T\oiii\/ and T\siii\/ ({\it top}), T\nii\/ ({\it middle}) and T\oii\/ ({\it bottom}).
The data points are compared with the model predictions of \citet[G92: red full line]{Garnett:1992}, 
\citet[P05: dashed blue line]{Perez-Montero:2005}, the high metallicity models of \citet[I06: dotted curve]{Izotov:2006}, and \citet[S90: dashed blue line]{Stasinska:1990}. The dashed lines in the bottom panel show models from \citet{Perez-Montero:2003}, calculated for \den\,=\,10, 100 and 500 cm\,$^{-3}$.\label{te-fig}}
\epsscale{1}
\end{figure}

In the analysis of extragalactic \hii\/ regions it is customary to assume a nebular structure that is described by a 
two- or a three-zone representation. Each zone is defined by an electron temperature, at which specific ions
are found to be the dominant emitting species. In this paper we assume a three-zone representation, characterized by the temperatures
T(\opp), T(\spp) and T(\op), in order of decreasing ionization potential of the ions that provide suitable emission
lines for the temperature measurement. We assume that T(\opp) is the temperature at which the \oiii\/ and \neiii\/ lines are emitted
(=\,T\oiii), and which is derived from the \oiii\lin4363/\llin4959,5007 line ratio. The ions \siii\/ and \ariii\/ are assumed to be found at the temperature T(\spp), which is measured as T\siii\/ from the \siii\lin6312/\llin9069,9532 
line ratio. Finally, the lower excitation ions \oii, \sii\/ and \nii\/ are found in the zone characterized by 
T(\op), which is typically measured as T\oii\/ from \oii\lin7325/\lin3727. Alternatively, one could also 
measure T(\op) from nitrogen or sulphur lines, using the \nii\lin5755/\llin6548,6583 and \sii\lin4072/\llin6717,6731 auroral-to-nebular line ratios.
Photoionization models predict correlations between the temperatures of the different zones (\citealt{Stasinska:1982,Stasinska:1990}; \citealt{Campbell:1986}; \citealt{Garnett:1992}; \citealt{Perez-Montero:2003}; \citealt{Izotov:2006}, \citealt{Pilyugin:2007b}). This is often exploited to infer the line temperature for ions that are not detected
in nebular spectra.

We have derived electron temperatures \te\/ from the different line ratios mentioned in the previous paragraph, using the {\it temden} program (\citealt{De-Robertis:1987}), implemented in {\sc iraf}'s {\it nebular} package (\citealt{Shaw:1995}).
We updated most of the atomic parameters used by the code, adopting collisional strengths and transition probabilites from recent sources in the literature, and summarized in Table~\ref{atomic} for the ions of interest.
The electron densities \den\/ presented in Col.~7 of Table~\ref{te} were derived iteratively as a function of \te\/ with {\it temden} from the \sii\lin6717/\sii\lin6731 line ratios. At very low densities (\lin6717/\lin6731 approaching its maximum theoretical limit) we
set \den\,=\,20\,cm\,$^{-3}$. 

In Fig.~\ref{te-fig} we plot the relationships between T\oiii\/ and T\siii\/ {\it (top)}, T\nii\/ {\it (middle)} and T\oii\/ {\it (bottom)}. We include for comparison the correlations predicted by photoionization models: \citet[G92]{Garnett:1992}, 
\citet[P05]{Perez-Montero:2005}, \citet[I06, their highest metallicity bin]{Izotov:2006}, and \citet[S90]{Stasinska:1990}, using for the latter the fits provided by \citet{Izotov:1994}.
The top panel  shows a tight correlation between T\oiii\/ and T\siii, confirming  earlier 
empirical measurements in \hii\/ regions within M101 by \citet{Kennicutt:2003}. The rms deviations of the data points from the model predictions of \citet[full red line]{Garnett:1992} and \citet[black dotted curve]{Izotov:2006} are comparable (410~K \vs~470~K), while the models by 
\citet{Perez-Montero:2005} provide a slightly worse fit to the data (the scatter around their model line is 560~K).

In the case of T\nii\/ \vs~T\oiii\/ (Fig.~\ref{te-fig}, {\it middle}) a clear correlation is also seen,
but the fit to the model predictions is not as good as in the T\siii\/ case. Formally the \citet{Izotov:2006} models provide the best fit to the observations, but the error bars for the points with the highest electron temperatures are large, while below 10$^4$~K the \citet{Izotov:2006} models are virtually indistinguishable from the \citet{Garnett:1992} models. 

\begin{deluxetable}{lcc}
\tabletypesize{\scriptsize}
\tabletypesize{\footnotesize}
\tablecolumns{3}
\tablewidth{0pt}
\tablecaption{Sources of atomic data\label{atomic}}

\tablehead{
\colhead{\phantom{}Ion\phantom{}}	     &
\colhead{\phantom{}Transition probabilities\phantom{}}       &
\colhead{\phantom{}Collision strengths\phantom{}}	}
\startdata
\\[-2mm]
\oii			& Froese Fischer \&								& \citet{Tayal:2007}\tablenotemark{a} 			\\
			& Tachiev (2004)\nocite{Froese-Fischer:2004}\tablenotemark{a}				&												\\[1mm]
			
\oiii		& Froese Fischer \&								& \citet{Aggarwal:1999}\tablenotemark{a} 			\\
			& Tachiev (2004)\tablenotemark{a}				&												\\[1mm]
			
\nii			& Froese Fischer \&								& \citet{Hudson:2005}\tablenotemark{a} 			\\
			& Tachiev (2004)\tablenotemark{a}				&												\\[1mm]

\neiii		& Froese Fischer \&								& \citet{McLaughlin:2000}\tablenotemark{a} 			\\
			& Tachiev (2004)\tablenotemark{a}				&												\\[1mm]

\ariii		& \citet{Mendoza:1983}		& \citet{Galavis:1995} 								\\
			& \citet{Kaufman:1986}		& 													\\[1mm]

\sii			& \citet{Froese-Fischer:2006}\tablenotemark{a}	& \citet{Ramsbottom:1996} 							\\[1mm]

\siii		& \citet{Froese-Fischer:2006}\tablenotemark{a}	& \citet{Tayal:1999}\tablenotemark{a} 				
			
\enddata 
\tablenotetext{a}{Updated from {\sc iraf}'s value.\\}
\end{deluxetable}

The bottom panel of Fig.~\ref{te-fig} shows that the observed T\oii\/ \vs~T\oiii\/ relation is in rough agreement with the \citet{Garnett:1992} prediction, although the scatter  is considerably 
larger than in the previous two cases. This could be related to the fact that 
the T\oii\/ temperature is sensitive to the electron density through collisional de-excitation. At 10$^4$~K a change of 50\,cm\,$^{-3}$ around a nominal density of 100\,cm\,$^{-3}$ results in a temperature variation of $\pm$350~K (the effect
is instead negligible for the other ion temperatures, except for T\sii, for which the variation is $\pm$260~K).  Predictions of the relation between T\oii\/ and T\oiii\/ from \citet{Perez-Montero:2003} are shown for three different values of \den: 10, 100 and 500 cm\,$^{-3}$. The density values we derive from \sii\lin6717/\sii\lin6731  are mostly below 100\,cm\,$^{-3}$ (only in four cases we measure
densities that are only moderately above this limit, up to 250\,cm\,$^{-3}$; see  Table~\ref{te}), and we do not see a correlation between the measured \den\/ and the position of the data points in the diagram. The  rms scatter of the points around the 
\citet{Garnett:1992} model prediction is 800~K, and does not increase with temperature, as instead suggested by the \citet{Perez-Montero:2003} models. \citet{Kennicutt:2003} list a number of physical reasons, besides collisional de-excitation, that could help explain the scatter seen in the relation between 
T\oii\/ and T\oiii, including shocks and radiative transfer
effects. Observational uncertainties might play a role, too, given the large wavelength baseline between \oii\lin3727 and the auroral line \oii\lin7325, and the contamination of the 
latter (which is actually  \oii\llin7320,7330) by airglow OH lines at \llin7316, 7329, 7341~\AA\/ (wavelengths from \citealt{Osterbrock:1996}). 
We note that we have corrected the auroral line intensity \oii\lin7325/\hbeta\/ for recombination into the upper level (following \citealt{Liu:2000}), although the effect is found to be at most 6\%.

\begin{figure}
\medskip
\center \includegraphics[width=0.47\textwidth]{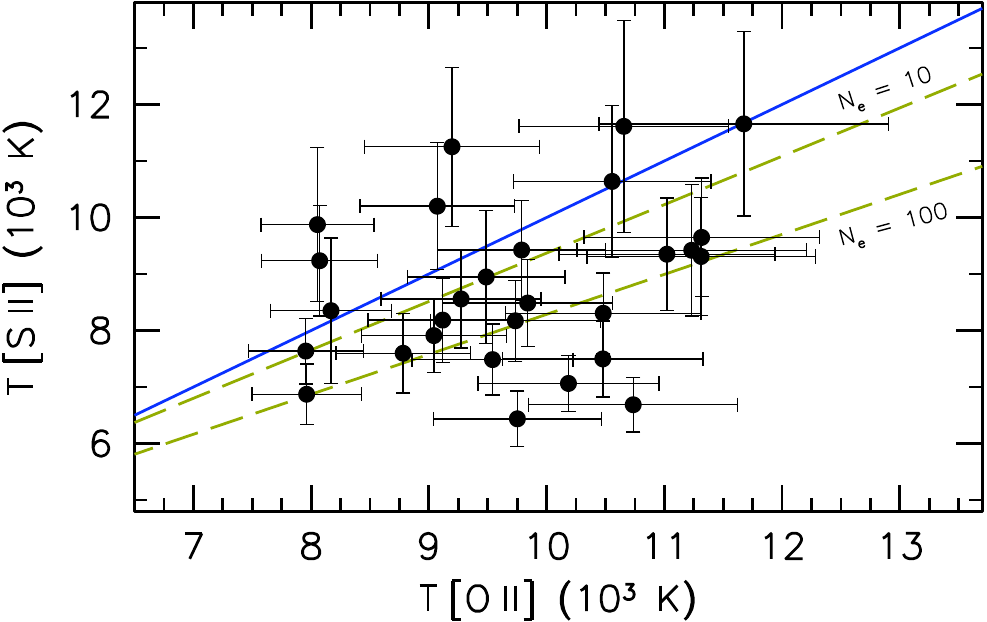}\medskip
\caption{Relation between T\oii\/ and T\sii\/. The dashed lines  show models from \citet{Perez-Montero:2003}, calculated for \den\,=\,10 and 100 cm\,$^{-3}$. The blue continuous line represents the one-to-one relation.\label{te2-fig}\\ }
\end{figure}

We have measured the \sii\llin4069,4076 auroral lines (abbreviated \sii\lin4072 in Table~\ref{fluxes_a}) for all the objects in our sample, allowing the derivation of T\sii\/ temperatures. In many cases, however, the measurement of these lines is made uncertain at the moderate resolution of our spectra by the presence of the nearby \hgamma\/ line, which often presents absorption wing components from the 
underlying stellar population. In Fig.~\ref{te2-fig}, which shows T\sii\/ as a function of T\oii, we do not observe any 
clear correlation. The plot agrees on average with the predictions by \citet{Perez-Montero:2003}, shown for \den\,=\,10 and 100 cm\,$^{-3}$, that T\sii\,$<$\,T\oii. This result could arise from the fact that the \sulp-emitting zone does not coincide with the \op\/ zone, due to the lower ionization potential of S$^0$ (10.4~eV) relative to that of O$^0$ (13.6~eV). The scatter of the points in Fig.~\ref{te2-fig}
is very large, and also in this case we do not find that the position in the diagram corresponds to the measured value of \den, i.e.~the points at the bottom do not 
have systematically larger densities.


\begin{deluxetable*}{ccccccc}
\tabletypesize{\scriptsize}
\tabletypesize{\footnotesize}
\tablecolumns{7}
\tablewidth{0pt}
\tablecaption{Measured electron temperatures (K) and densities (cm$^{-3}$)\label{te}}

\tablehead{
\colhead{\phantom{aaaaa}ID\phantom{aaaaa}}	     &
\colhead{\phantom{aaa}T\oiii\phantom{aaa}}       &
\colhead{\phantom{aaa}T\siii\phantom{aaa}}       &
\colhead{\phantom{a}T\nii\phantom{a}}	 			&
\colhead{\phantom{a}T\oii\phantom{a}}	 			&
\colhead{\phantom{a}T\sii\phantom{a}}	 			&
\colhead{\phantom{a}N$\rm_e$\phantom{a}}        \\[1mm]
\colhead{(1)}	&
\colhead{(2)}	&
\colhead{(3)}	&
\colhead{(4)}	&
\colhead{(5)}	&
\colhead{(6)}	&
\colhead{(7)}   }
\startdata
\\[-2mm]
 1\dotfill & 11700 $\pm$  500 &     \nodata        &     \nodata        &   10700 $\pm$  900 &   11600 $\pm$ 1900 &     253     \\ 
 2\dotfill & 11900 $\pm$  500 &   11800 $\pm$  900 &   12600 $\pm$ 1600 &   11300 $\pm$ 1000 &    9600 $\pm$ 1000 &     182     \\ 
 3\dotfill & 11400 $\pm$  900 &     \nodata        &     \nodata        &   11200 $\pm$ 1000 &    9400 $\pm$ 1200 &      20     \\ 
 4\dotfill & 10900 $\pm$  500 &   11100 $\pm$  900 &     \nodata        &   11000 $\pm$  900 &    9300 $\pm$ 1000 &      25     \\ 
 5\dotfill &  9200 $\pm$  500 &    9000 $\pm$  600 &     \nodata        &    9500 $\pm$  700 &    8900 $\pm$ 1200 &      20     \\ 
 6\dotfill &  9100 $\pm$  500 &   10100 $\pm$  800 &     \nodata        &    9200 $\pm$  700 &   11300 $\pm$ 1400 &      20     \\ 
 7\dotfill & 11500 $\pm$  500 &     \nodata        &     \nodata        &   11700 $\pm$ 1200 &   11700 $\pm$ 1600 &     247     \\ 
 8\dotfill &  9400 $\pm$  500 &    9900 $\pm$  700 &   10200 $\pm$ 1000 &    9000 $\pm$  600 &    7900 $\pm$  700 &      20     \\ 
 9\dotfill &  8600 $\pm$  300 &    8200 $\pm$  400 &     \nodata        &    9700 $\pm$  700 &    8200 $\pm$  700 &     115     \\ 
10\dotfill &  8600 $\pm$  300 &    8300 $\pm$  400 &    9300 $\pm$  600 &    8100 $\pm$  500 &    9200 $\pm$ 1000 &      91     \\ 
11\dotfill &  9500 $\pm$  600 &    9300 $\pm$  600 &     \nodata        &   11300 $\pm$ 1000 &    9300 $\pm$ 1000 &      20     \\ 
12\dotfill &   \nodata        &    7700 $\pm$  400 &    7800 $\pm$  800 &    8000 $\pm$  500 &    6900 $\pm$  500 &      16     \\ 
13\dotfill &   \nodata        &    8800 $\pm$  500 &   10500 $\pm$  900 &    8200 $\pm$  500 &    8400 $\pm$ 1300 &      14     \\ 
14\dotfill &  8800 $\pm$  300 &    8700 $\pm$  400 &    8700 $\pm$  500 &   10200 $\pm$  800 &    7100 $\pm$  500 &      57     \\ 
15\dotfill &   \nodata        &    8100 $\pm$  400 &    8700 $\pm$  800 &    9500 $\pm$  700 &    7500 $\pm$  600 &      30     \\ 
16\dotfill &   \nodata        &    8200 $\pm$  600 &    9600 $\pm$ 1000 &    9100 $\pm$  700 &   10200 $\pm$ 1100 &      20     \\ 
17\dotfill &  8100 $\pm$  300 &    8800 $\pm$  500 &    9500 $\pm$  600 &    8000 $\pm$  500 &    7600 $\pm$  600 &      20     \\ 
18\dotfill &   \nodata        &    8800 $\pm$  500 &    8400 $\pm$  600 &   10700 $\pm$  900 &    6700 $\pm$  500 &      20     \\ 
19\dotfill &  8500 $\pm$  300 &    8300 $\pm$  400 &    9100 $\pm$  600 &    8100 $\pm$  500 &    9900 $\pm$ 1400 &      20     \\ 
20\dotfill &  8200 $\pm$  300 &    8200 $\pm$  400 &    9400 $\pm$  600 &    9100 $\pm$  600 &    8200 $\pm$  700 &      91     \\ 
21\dotfill &   \nodata        &    8000 $\pm$  500 &    8400 $\pm$  600 &    9300 $\pm$  700 &    8600 $\pm$  900 &      87     \\ 
22\dotfill &   \nodata        &    7900 $\pm$  400 &    8000 $\pm$  500 &    8800 $\pm$  600 &    7600 $\pm$  700 &      34     \\ 
23\dotfill &  8000 $\pm$  400 &    8300 $\pm$  500 &    8800 $\pm$  500 &    9800 $\pm$  700 &    8500 $\pm$  800 &      30     \\ 
24\dotfill &  8400 $\pm$  300 &     \nodata        &    9200 $\pm$  500 &    9800 $\pm$  700 &    6400 $\pm$  500 &      88     \\ 
25\dotfill &   \nodata        &    8400 $\pm$  600 &    9900 $\pm$ 1000 &   10500 $\pm$  900 &    7500 $\pm$  700 &      54     \\ 
26\dotfill &  9000 $\pm$  400 &    9000 $\pm$  500 &    9800 $\pm$  600 &   10600 $\pm$  800 &   10600 $\pm$ 1300 &      43     \\ 
27\dotfill & 11100 $\pm$  500 &   11300 $\pm$  800 &   10900 $\pm$  900 &    9800 $\pm$  700 &    9400 $\pm$  900 &      20     \\ 
28\dotfill & 10500 $\pm$  400 &   10300 $\pm$  900 &   11800 $\pm$ 1700 &   10500 $\pm$  800 &    8300 $\pm$  700 &      20     \\ 
\\[-6mm]
\enddata
\end{deluxetable*}

\begin{deluxetable}{cccc}
\tabletypesize{\scriptsize}
\tabletypesize{\footnotesize}
\tablecolumns{4}
\tablecaption{Adopted electron temperatures (K)\label{te2}}

\tablehead{
\colhead{\phantom{aaa}ID\phantom{aaa}}	     &
\colhead{\phantom{aa}T(\opp, \nepp)\phantom{aa}}       &
\colhead{\phantom{aa}T(\spp, \arpp)\phantom{aa}}       &
\colhead{\phantom{a}T(\op, \np, \sulp)\phantom{a}}	 \\[1mm]
\colhead{(1)}	&
\colhead{(2)}	&
\colhead{(3)}	&
\colhead{(4)}	}
\startdata
\\[-2mm]
 1\dotfill & 11700 $\pm$  500 &   11400 $\pm$  500 &   11200 $\pm$  600    \\ 
 2\dotfill & 12000 $\pm$  500 &   11600 $\pm$  400 &   11400 $\pm$  500    \\ 
 3\dotfill & 11400 $\pm$  900 &   11100 $\pm$  800 &   11000 $\pm$  700    \\ 
 4\dotfill & 11000 $\pm$  400 &   10800 $\pm$  400 &   10700 $\pm$  500    \\ 
 5\dotfill &  9100 $\pm$  400 &    9200 $\pm$  400 &    9300 $\pm$  500    \\ 
 6\dotfill &  9300 $\pm$  400 &    9400 $\pm$  400 &    9500 $\pm$  500    \\ 
 7\dotfill & 11500 $\pm$  500 &   11200 $\pm$  500 &   11000 $\pm$  500    \\ 
 8\dotfill &  9500 $\pm$  400 &    9600 $\pm$  400 &    9700 $\pm$  500    \\ 
 9\dotfill &  8300 $\pm$  300 &    8600 $\pm$  300 &    8800 $\pm$  400    \\ 
10\dotfill &  8400 $\pm$  300 &    8600 $\pm$  300 &    8900 $\pm$  400    \\ 
11\dotfill &  9400 $\pm$  500 &    9500 $\pm$  400 &    9600 $\pm$  500    \\ 
12\dotfill &  7200 $\pm$  500 &    7700 $\pm$  400 &    8000 $\pm$  600    \\ 
13\dotfill &  8600 $\pm$  700 &    8800 $\pm$  500 &    9000 $\pm$  600    \\ 
14\dotfill &  8700 $\pm$  300 &    8900 $\pm$  300 &    9100 $\pm$  400    \\ 
15\dotfill &  7700 $\pm$  600 &    8100 $\pm$  400 &    8400 $\pm$  600    \\ 
16\dotfill &  7900 $\pm$  700 &    8200 $\pm$  600 &    8500 $\pm$  600    \\ 
17\dotfill &  8200 $\pm$  300 &    8600 $\pm$  300 &    8700 $\pm$  400    \\ 
18\dotfill &  8600 $\pm$  600 &    8800 $\pm$  500 &    9000 $\pm$  600    \\ 
19\dotfill &  8400 $\pm$  300 &    8600 $\pm$  300 &    8900 $\pm$  400    \\ 
20\dotfill &  8100 $\pm$  200 &    8400 $\pm$  200 &    8700 $\pm$  400    \\ 
21\dotfill &  7600 $\pm$  600 &    8000 $\pm$  500 &    8300 $\pm$  600    \\ 
22\dotfill &  7500 $\pm$  600 &    7900 $\pm$  400 &    8300 $\pm$  600    \\ 
23\dotfill &  8000 $\pm$  300 &    8300 $\pm$  300 &    8600 $\pm$  500    \\ 
24\dotfill &  8400 $\pm$  300 &    8700 $\pm$  300 &    8900 $\pm$  500    \\ 
25\dotfill &  8100 $\pm$  700 &    8400 $\pm$  600 &    8700 $\pm$  700    \\ 
26\dotfill &  9000 $\pm$  300 &    9100 $\pm$  300 &    9300 $\pm$  500    \\ 
27\dotfill & 11200 $\pm$  400 &   11000 $\pm$  400 &   10800 $\pm$  500    \\ 
28\dotfill & 10500 $\pm$  400 &   10400 $\pm$  400 &   10400 $\pm$  500    \\ 
\\[-6mm]
\enddata
\end{deluxetable}

In light of the results obtained for the different ion temperatures, and in particular of the tightness of the relation between T\oiii\/ and T\siii\/ and the good match with the model predictions, we have adopted a three-zone ionization representation of the \hii\/ regions in NGC~300, and, following \citet{Kennicutt:2003} and \citet{Bresolin:2004, Bresolin:2005}, we used the relations between temperatures in the different zones given by
\citet{Garnett:1992}:

\begin{equation}\label{siiieqn}
\rm T[S\,III]  = 0.83\,T[O\,III]\, +\, 1700~ K
\end{equation}

\begin{equation}\label{niieqn}
\rm T[N\,II] = T[O\,II] = 0.70\,T[O\,III]\, +\, 3000~ K.
\end{equation}

\noindent
These scaling relations allow us to calculate the temperatures in the intermediate- and low-ionization zones, once the high-ionization zone temperature T\oiii\/ is known. On the other hand, in the eight \hii\/ regions of our sample for which \oiii\lin4363 was not detected, it is possible to derive T\oiii\/ from the knowledge of T\siii\/ through Eq.~(\ref{siiieqn}). In order to reduce the random errors when both T\oiii\/ and T\siii\/ are available, we adopted for the temperature of the high-ionization zone (where the emission from \opp\/ and \nepp\/ originates)
the 
weighted average between T\oiii\/ and the temperature derived by the inversion of Eq.~(\ref{siiieqn}).
In a similar way, for the intermediate-ionization zone (\spp, \arpp) we calculated a weighted average of T\siii\/ and the temperature resulting from the scaling relation. For the low-ionization zone (\op, \np, \sulp) we took the T\oii\/ value that results from
the insertion of the high-ionization zone temperature derived in the previous step into Eq.~(\ref{niieqn}). The temperatures for the three ionization zones derived with this method are summarized in Table~\ref{te2}. The errors quoted 
reflect the 1$\sigma$ uncertainties in the line fluxes used for the computation of the temperatures, and include an additional term, added in quadrature, that estimates the uncertainty in the scaling relations ($\pm$200\,K for
T\siii\/ \vs~T\oiii, $\pm$400\,K for T\oii\/ \vs~T\oiii).

\begin{figure}
\medskip
\center \includegraphics[width=0.47\textwidth]{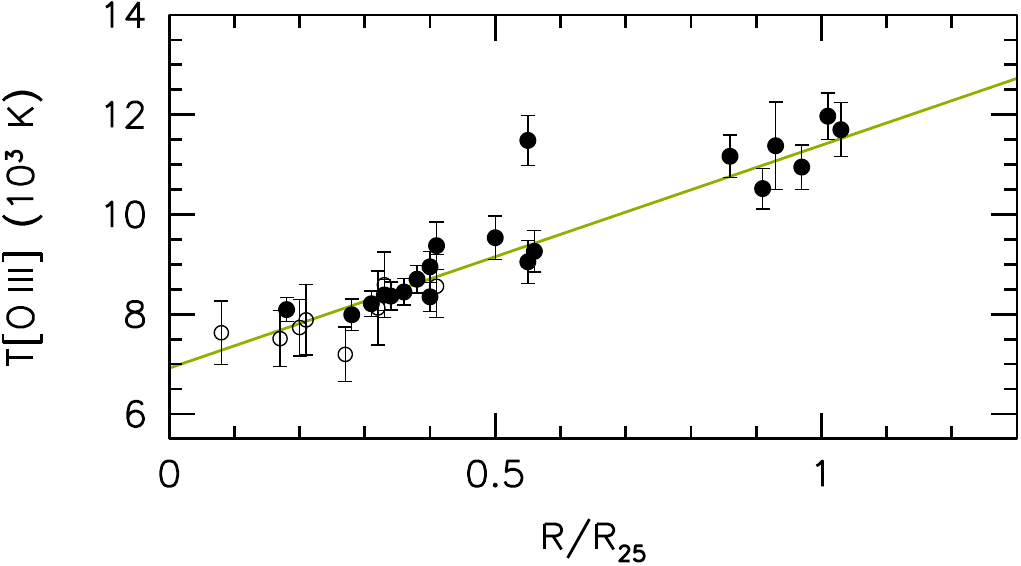}\medskip
\caption{The T\oiii\/ radial gradient, in terms of the isophotal radius \rtf. The linear least-square fit to the data, with the exception of the 
outlier (region \#7\,=\,De\,30), is shown by the line. In this and several of the following plots open disk symbols are used to represent the eight \hii\/ regions whose \oiii\lin4363 line is not detected, and for which \te\/ was derived from \siii\lin6312 alone.\label{fig_ter}\\ }
\epsscale{1}
\end{figure}

In Fig.~\ref{fig_ter} we plot the radial trend of T\oiii, where we have used full symbols for the \hii\/ regions for which we determined this temperature directly from \oiii\lin4363, and open symbols for the eight \hii\/ regions for which we derived T\oiii\/ from \siii\lin6312 and inverting equation~(\ref{siiieqn}).
There is a clear outlier, object \#7 in our list (=\,De\,30, further discussed in \S~\ref{wrsect} and \S~\ref{ionizing}), with respect to the temperature gradient defined by the remaining \hii\/ regions. After excluding this object, a weighted linear least-square fit yields:

\begin{equation}
{\rm T[O\,III] = 6920~(\pm 130)\ +\ 4470~(\pm 260)}~R/R_{25}
\end{equation}
\noindent
where the temperature is expressed in degrees K. The regression is shown with a continuous line in Fig.~\ref{fig_ter}. The gradient corresponds to $840\pm50$~K\,kpc$^{-1}$. This value can be compared to the results obtained for other galaxies, such as M33, where \citet{Magrini:2007} measured $570\pm130$~K\,kpc$^{-1}$,  and the Milky Way, for which \citet{Deharveng:2000} found $372\pm38$~K\,kpc$^{-1}$ from hydrogen radio recombination lines (a flatter slope, $287\pm46$~K\,kpc$^{-1}$, was obtained by \citealt{Quireza:2006}). For M101, from \citet{Kennicutt:2003} we derived $290\pm25$~K\,kpc$^{-1}$.
This quick comparison suggests that smaller, less luminous spiral galaxies  have steeper temperature gradients than larger ones. Since the temperature gradients are due to variations in abundances, with cooling via
line emission being more effective at high metallicities, this is a reflection of the fact that more luminous spiral galaxies tend to have flatter abundance gradients when expressed in dex\,kpc$^{-1}$ (\citealt{Garnett:1997}).


\section{Chemical abundances}
The ionic abundances were derived with the {\it ionic} program in {\sc iraf} from the electron temperatures adopted in the three ionization zones (Table~ \ref{te2}) and the reddening-corrected emission line fluxes in Tables~\ref{fluxes_a} and \ref{fluxes_b}. The results are presented in Table~\ref{ionic}, where for each ionic species $X$ we provide the quantity
12\,+\,log($X$/H$^+$).

\begin{deluxetable*}{cccccccc}
\tabletypesize{\scriptsize}
\tablecolumns{8}
\tablewidth{0pt}
\tablecaption{Ionic abundances: 12\,+\,log($X$/H$^+$)\label{ionic}}

\tablehead{
\colhead{\phantom{aaaaa}ID\phantom{aaaaa}}	     &
\colhead{\phantom{}\op\phantom{}}       &
\colhead{\phantom{}\opp\phantom{}}       &
\colhead{\phantom{}\sulp\phantom{}}       &
\colhead{\phantom{}\spp\phantom{}}       &
\colhead{\phantom{}\np\phantom{}}       &
\colhead{\phantom{}\nepp\phantom{}}       &
\colhead{\phantom{}\arpp\phantom{}}	 \\[1mm]
\colhead{(1)}	&
\colhead{(2)}	&
\colhead{(3)}	&
\colhead{(4)}	&
\colhead{(5)}	&
\colhead{(6)}	&
\colhead{(7)}	&
\colhead{(8)}	}
\startdata
\\[-2mm]
 1\dotfill &  7.78 $\pm$ 0.09 &    7.85 $\pm$ 0.06 &    5.62 $\pm$ 0.05 &    6.31 $\pm$ 0.03 &    6.29 $\pm$ 0.05 &    7.10 $\pm$ 0.07 &    5.70 $\pm$ 0.04    \\ 
 2\dotfill &  7.58 $\pm$ 0.08 &    8.02 $\pm$ 0.05 &    5.49 $\pm$ 0.04 &    6.36 $\pm$ 0.03 &    6.03 $\pm$ 0.05 &    7.27 $\pm$ 0.06 &    5.75 $\pm$ 0.03    \\ 
 3\dotfill &  7.98 $\pm$ 0.12 &    7.46 $\pm$ 0.11 &    5.98 $\pm$ 0.07 &    6.21 $\pm$ 0.06 &    6.55 $\pm$ 0.07 &    6.49 $\pm$ 0.12 &    5.59 $\pm$ 0.06    \\ 
 4\dotfill &  7.91 $\pm$ 0.09 &    7.84 $\pm$ 0.06 &    5.81 $\pm$ 0.05 &    6.44 $\pm$ 0.03 &    6.52 $\pm$ 0.05 &    7.06 $\pm$ 0.07 &    5.85 $\pm$ 0.04    \\ 
 5\dotfill &  8.18 $\pm$ 0.11 &    8.08 $\pm$ 0.08 &    6.16 $\pm$ 0.06 &    6.69 $\pm$ 0.04 &    6.84 $\pm$ 0.07 &    7.24 $\pm$ 0.09 &    6.00 $\pm$ 0.05    \\ 
 6\dotfill &  8.11 $\pm$ 0.10 &    8.08 $\pm$ 0.07 &    6.00 $\pm$ 0.06 &    6.56 $\pm$ 0.04 &    6.78 $\pm$ 0.06 &    7.33 $\pm$ 0.09 &    5.97 $\pm$ 0.04    \\ 
 7\dotfill &  7.54 $\pm$ 0.08 &    8.33 $\pm$ 0.06 &    5.83 $\pm$ 0.05 &    6.58 $\pm$ 0.03 &    6.36 $\pm$ 0.05 &    7.69 $\pm$ 0.07 &    6.02 $\pm$ 0.04    \\ 
 8\dotfill &  8.12 $\pm$ 0.10 &    7.80 $\pm$ 0.07 &    6.04 $\pm$ 0.06 &    6.54 $\pm$ 0.04 &    6.85 $\pm$ 0.06 &    6.99 $\pm$ 0.09 &    5.90 $\pm$ 0.04    \\ 
 9\dotfill &  8.06 $\pm$ 0.11 &    8.23 $\pm$ 0.06 &    6.03 $\pm$ 0.06 &    6.77 $\pm$ 0.03 &    6.94 $\pm$ 0.07 &    7.47 $\pm$ 0.07 &    6.15 $\pm$ 0.04    \\ 
10\dotfill &  8.06 $\pm$ 0.11 &    8.21 $\pm$ 0.05 &    5.71 $\pm$ 0.06 &    6.75 $\pm$ 0.03 &    6.70 $\pm$ 0.06 &    7.49 $\pm$ 0.07 &    6.10 $\pm$ 0.03    \\ 
11\dotfill &  8.07 $\pm$ 0.11 &    7.98 $\pm$ 0.08 &    6.03 $\pm$ 0.06 &    6.62 $\pm$ 0.04 &    6.80 $\pm$ 0.07 &    7.17 $\pm$ 0.10 &    6.05 $\pm$ 0.05    \\ 
12\dotfill &  8.44 $\pm$ 0.16 &    7.62 $\pm$ 0.16 &    6.47 $\pm$ 0.09 &    6.80 $\pm$ 0.06 &    7.29 $\pm$ 0.10 &    0.00 $\pm$ 0.00 &    5.93 $\pm$ 0.07    \\ 
13\dotfill &  8.17 $\pm$ 0.14 &    7.89 $\pm$ 0.13 &    6.05 $\pm$ 0.08 &    6.64 $\pm$ 0.06 &    7.01 $\pm$ 0.09 &    6.92 $\pm$ 0.16 &    5.93 $\pm$ 0.07    \\ 
14\dotfill &  8.16 $\pm$ 0.10 &    8.05 $\pm$ 0.05 &    6.09 $\pm$ 0.06 &    6.77 $\pm$ 0.03 &    7.01 $\pm$ 0.06 &    7.22 $\pm$ 0.06 &    6.17 $\pm$ 0.03    \\ 
15\dotfill &  8.19 $\pm$ 0.15 &    7.97 $\pm$ 0.14 &    6.24 $\pm$ 0.08 &    6.84 $\pm$ 0.06 &    7.26 $\pm$ 0.09 &    6.98 $\pm$ 0.17 &    6.15 $\pm$ 0.07    \\ 
16\dotfill &  8.31 $\pm$ 0.17 &    7.94 $\pm$ 0.17 &    6.26 $\pm$ 0.09 &    6.76 $\pm$ 0.07 &    7.21 $\pm$ 0.10 &    7.05 $\pm$ 0.20 &    6.10 $\pm$ 0.08    \\ 
17\dotfill &  8.17 $\pm$ 0.11 &    8.18 $\pm$ 0.06 &    5.94 $\pm$ 0.06 &    6.78 $\pm$ 0.03 &    6.89 $\pm$ 0.07 &    7.25 $\pm$ 0.07 &    6.08 $\pm$ 0.04    \\ 
18\dotfill &  8.19 $\pm$ 0.14 &    7.89 $\pm$ 0.13 &    6.23 $\pm$ 0.08 &    6.73 $\pm$ 0.06 &    7.08 $\pm$ 0.09 &    7.01 $\pm$ 0.15 &    6.13 $\pm$ 0.06    \\ 
19\dotfill &  8.09 $\pm$ 0.11 &    8.08 $\pm$ 0.06 &    5.98 $\pm$ 0.06 &    6.70 $\pm$ 0.03 &    6.80 $\pm$ 0.06 &    7.22 $\pm$ 0.07 &    5.93 $\pm$ 0.04    \\ 
20\dotfill &  8.03 $\pm$ 0.11 &    8.27 $\pm$ 0.05 &    5.83 $\pm$ 0.06 &    6.82 $\pm$ 0.03 &    6.83 $\pm$ 0.07 &    7.46 $\pm$ 0.07 &    6.19 $\pm$ 0.03    \\ 
21\dotfill &  8.28 $\pm$ 0.17 &    8.04 $\pm$ 0.17 &    6.32 $\pm$ 0.09 &    6.86 $\pm$ 0.07 &    7.37 $\pm$ 0.10 &    7.12 $\pm$ 0.19 &    6.22 $\pm$ 0.08    \\ 
22\dotfill &  8.39 $\pm$ 0.16 &    7.94 $\pm$ 0.15 &    6.44 $\pm$ 0.09 &    6.77 $\pm$ 0.06 &    7.31 $\pm$ 0.09 &    7.15 $\pm$ 0.17 &    6.11 $\pm$ 0.07    \\ 
23\dotfill &  8.12 $\pm$ 0.12 &    8.24 $\pm$ 0.07 &    5.98 $\pm$ 0.07 &    6.75 $\pm$ 0.03 &    6.91 $\pm$ 0.07 &    7.48 $\pm$ 0.09 &    6.19 $\pm$ 0.04    \\ 
24\dotfill &  8.11 $\pm$ 0.11 &    8.11 $\pm$ 0.06 &    5.99 $\pm$ 0.06 &    6.75 $\pm$ 0.04 &    6.94 $\pm$ 0.07 &    7.10 $\pm$ 0.08 &    6.09 $\pm$ 0.04    \\ 
25\dotfill &  8.31 $\pm$ 0.17 &    8.04 $\pm$ 0.17 &    6.18 $\pm$ 0.09 &    6.74 $\pm$ 0.07 &    7.10 $\pm$ 0.10 &    7.08 $\pm$ 0.20 &    6.18 $\pm$ 0.08    \\ 
26\dotfill &  7.92 $\pm$ 0.10 &    8.16 $\pm$ 0.06 &    5.79 $\pm$ 0.06 &    6.73 $\pm$ 0.03 &    6.68 $\pm$ 0.06 &    7.30 $\pm$ 0.07 &    6.12 $\pm$ 0.04    \\ 
27\dotfill &  7.98 $\pm$ 0.08 &    7.66 $\pm$ 0.05 &    6.19 $\pm$ 0.05 &    6.35 $\pm$ 0.03 &    6.70 $\pm$ 0.05 &    6.85 $\pm$ 0.06 &    5.67 $\pm$ 0.03    \\ 
28\dotfill &  8.00 $\pm$ 0.09 &    7.88 $\pm$ 0.06 &    5.89 $\pm$ 0.05 &    6.49 $\pm$ 0.03 &    6.57 $\pm$ 0.05 &    7.06 $\pm$ 0.07 &    5.82 $\pm$ 0.04    \\ 
\enddata
\\[-6mm]
\end{deluxetable*}

\begin{deluxetable*}{cccccccc}
\tabletypesize{\scriptsize}
\tablecolumns{8}
\tablewidth{0pt}
\tablecaption{Total abundances\label{abundances}}

\tablehead{
\colhead{\phantom{aaaaa}ID\phantom{aaaaa}}	     &
\colhead{\phantom{}R/R$_{25}$\phantom{}}       &
\colhead{\phantom{}12\,+\,log\,(O/H)\phantom{}}       &
\colhead{\phantom{}log\,(N/O)\phantom{}}       &
\colhead{\phantom{}log\,(S/O)\phantom{}}       &
\colhead{\phantom{}log\,(Ar/O)\phantom{}}       &
\colhead{\phantom{}log\,(Ne/O)\phantom{}}	       &
\colhead{\phantom{}He/H\phantom{}}\\[1mm]
\colhead{(1)}	&
\colhead{(2)}	&
\colhead{(3)}	&
\colhead{(4)}	&
\colhead{(5)}	&
\colhead{(6)}	&
\colhead{(7)}	&
\colhead{(8)}	}
\startdata
\\[-2mm]
 1\dotfill &   1.03  &  8.12 $\pm$ 0.05 &   $-1.49$ $\pm$ 0.10 &   $-1.70$ $\pm$ 0.06 &   $-2.39$ $\pm$ 0.08 &   $-0.76$ $\pm$ 0.10 &   0.080 $\pm$ 0.007    \\ 
 2\dotfill &   1.01  &  8.15 $\pm$ 0.04 &   $-1.55$ $\pm$ 0.09 &   $-1.69$ $\pm$ 0.05 &   $-2.36$ $\pm$ 0.07 &   $-0.75$ $\pm$ 0.08 &   0.092 $\pm$ 0.011    \\ 
 3\dotfill &   0.93  &  8.09 $\pm$ 0.09 &   $-1.43$ $\pm$ 0.14 &   $-1.66$ $\pm$ 0.10 &   $-2.46$ $\pm$ 0.13 &   $-0.98$ $\pm$ 0.17 &   0.083 $\pm$ 0.011    \\ 
 4\dotfill &   0.97  &  8.18 $\pm$ 0.05 &   $-1.39$ $\pm$ 0.10 &   $-1.62$ $\pm$ 0.06 &   $-2.30$ $\pm$ 0.07 &   $-0.78$ $\pm$ 0.09 &   0.102 $\pm$ 0.013    \\ 
 5\dotfill &   0.55  &  8.43 $\pm$ 0.07 &   $-1.34$ $\pm$ 0.13 &   $-1.63$ $\pm$ 0.08 &   $-2.38$ $\pm$ 0.10 &   $-0.84$ $\pm$ 0.12 &   0.086 $\pm$ 0.011    \\ 
 6\dotfill &   0.56  &  8.40 $\pm$ 0.06 &   $-1.33$ $\pm$ 0.12 &   $-1.74$ $\pm$ 0.07 &   $-2.38$ $\pm$ 0.09 &   $-0.75$ $\pm$ 0.11 &   0.092 $\pm$ 0.012    \\ 
 7\dotfill &   0.55  &  8.40 $\pm$ 0.05 &   $-1.18$ $\pm$ 0.10 &   $-1.50$ $\pm$ 0.06 &   $-2.32$ $\pm$ 0.08 &   $-0.64$ $\pm$ 0.09 &   0.083 $\pm$ 0.009    \\ 
 8\dotfill &   0.50  &  8.29 $\pm$ 0.07 &   $-1.27$ $\pm$ 0.12 &   $-1.65$ $\pm$ 0.08 &   $-2.32$ $\pm$ 0.09 &   $-0.81$ $\pm$ 0.11 &   0.085 $\pm$ 0.011    \\ 
 9\dotfill &   0.40  &  8.46 $\pm$ 0.06 &   $-1.12$ $\pm$ 0.13 &   $-1.58$ $\pm$ 0.06 &   $-2.28$ $\pm$ 0.08 &   $-0.76$ $\pm$ 0.10 &   0.092 $\pm$ 0.011    \\ 
10\dotfill &   0.36  &  8.45 $\pm$ 0.05 &   $-1.36$ $\pm$ 0.12 &   $-1.63$ $\pm$ 0.06 &   $-2.31$ $\pm$ 0.07 &   $-0.72$ $\pm$ 0.09 &   0.079 $\pm$ 0.010    \\ 
11\dotfill &   0.41  &  8.33 $\pm$ 0.07 &   $-1.27$ $\pm$ 0.13 &   $-1.61$ $\pm$ 0.08 &   $-2.22$ $\pm$ 0.10 &   $-0.81$ $\pm$ 0.13 &   0.089 $\pm$ 0.011    \\ 
12\dotfill &   0.27  &  8.50 $\pm$ 0.14 &   $-1.15$ $\pm$ 0.19 &   $-1.56$ $\pm$ 0.15 &   $-2.47$ $\pm$ 0.17 &   \nodata &0.060 $\pm$ 0.007    \\ 
13\dotfill &   0.33  &  8.35 $\pm$ 0.11 &   $-1.15$ $\pm$ 0.17 &   $-1.63$ $\pm$ 0.12 &   $-2.36$ $\pm$ 0.15 &   $-0.97$ $\pm$ 0.21 &   0.083 $\pm$ 0.010    \\ 
14\dotfill &   0.38  &  8.41 $\pm$ 0.06 &   $-1.15$ $\pm$ 0.12 &   $-1.56$ $\pm$ 0.07 &   $-2.18$ $\pm$ 0.07 &   $-0.83$ $\pm$ 0.08 &   0.091 $\pm$ 0.011    \\ 
15\dotfill &   0.20  &  8.39 $\pm$ 0.11 &   $-0.93$ $\pm$ 0.18 &   $-1.47$ $\pm$ 0.12 &   $-2.18$ $\pm$ 0.16 &   $-0.99$ $\pm$ 0.23 &   0.088 $\pm$ 0.011    \\ 
16\dotfill &   0.21  &  8.47 $\pm$ 0.13 &   $-1.10$ $\pm$ 0.20 &   $-1.61$ $\pm$ 0.14 &   $-2.30$ $\pm$ 0.19 &   $-0.89$ $\pm$ 0.28 &   0.079 $\pm$ 0.010    \\ 
17\dotfill &   0.31  &  8.47 $\pm$ 0.06 &   $-1.28$ $\pm$ 0.13 &   $-1.63$ $\pm$ 0.07 &   $-2.35$ $\pm$ 0.07 &   $-0.92$ $\pm$ 0.09 &   0.091 $\pm$ 0.011    \\ 
18\dotfill &   0.41  &  8.37 $\pm$ 0.10 &   $-1.11$ $\pm$ 0.17 &   $-1.54$ $\pm$ 0.11 &   $-2.17$ $\pm$ 0.15 &   $-0.88$ $\pm$ 0.21 &   0.090 $\pm$ 0.011    \\ 
19\dotfill &   0.34  &  8.39 $\pm$ 0.06 &   $-1.30$ $\pm$ 0.13 &   $-1.61$ $\pm$ 0.07 &   $-2.41$ $\pm$ 0.08 &   $-0.85$ $\pm$ 0.09 &   0.103 $\pm$ 0.013    \\ 
20\dotfill &   0.18  &  8.47 $\pm$ 0.05 &   $-1.20$ $\pm$ 0.13 &   $-1.57$ $\pm$ 0.06 &   $-2.25$ $\pm$ 0.07 &   $-0.82$ $\pm$ 0.09 &   0.091 $\pm$ 0.011    \\ 
21\dotfill &   0.08  &  8.48 $\pm$ 0.12 &   $-0.92$ $\pm$ 0.19 &   $-1.52$ $\pm$ 0.13 &   $-2.19$ $\pm$ 0.19 &   $-0.92$ $\pm$ 0.26 &   0.075 $\pm$ 0.010    \\ 
22\dotfill &   0.17  &  8.53 $\pm$ 0.12 &   $-1.09$ $\pm$ 0.18 &   $-1.62$ $\pm$ 0.13 &   $-2.34$ $\pm$ 0.17 &   $-0.80$ $\pm$ 0.23 &   0.077 $\pm$ 0.009    \\ 
23\dotfill &   0.28  &  8.48 $\pm$ 0.06 &   $-1.21$ $\pm$ 0.14 &   $-1.65$ $\pm$ 0.07 &   $-2.26$ $\pm$ 0.09 &   $-0.75$ $\pm$ 0.11 &   0.085 $\pm$ 0.010    \\ 
24\dotfill &   0.33  &  8.41 $\pm$ 0.06 &   $-1.17$ $\pm$ 0.13 &   $-1.59$ $\pm$ 0.07 &   $-2.28$ $\pm$ 0.08 &   $-1.01$ $\pm$ 0.10 &   0.082 $\pm$ 0.010    \\ 
25\dotfill &   0.32  &  8.50 $\pm$ 0.13 &   $-1.22$ $\pm$ 0.20 &   $-1.67$ $\pm$ 0.14 &   $-2.25$ $\pm$ 0.19 &   $-0.96$ $\pm$ 0.28 &   0.081 $\pm$ 0.010    \\ 
26\dotfill &   0.40  &  8.36 $\pm$ 0.05 &   $-1.25$ $\pm$ 0.12 &   $-1.54$ $\pm$ 0.06 &   $-2.21$ $\pm$ 0.07 &   $-0.85$ $\pm$ 0.09 &   0.092 $\pm$ 0.011    \\ 
27\dotfill &   0.86  &  8.15 $\pm$ 0.06 &   $-1.29$ $\pm$ 0.10 &   $-1.55$ $\pm$ 0.06 &   $-2.44$ $\pm$ 0.07 &   $-0.82$ $\pm$ 0.08 &   0.084 $\pm$ 0.010    \\ 
28\dotfill &   0.91  &  8.25 $\pm$ 0.06 &   $-1.43$ $\pm$ 0.10 &   $-1.67$ $\pm$ 0.06 &   $-2.38$ $\pm$ 0.07 &   $-0.83$ $\pm$ 0.09 &   0.102 $\pm$ 0.012    \\ 
\\[-6mm]
\enddata
\end{deluxetable*}

In order to derive the total chemical abundances we need to adopt an ionization correction scheme, to account for unseen ionization stages of the chemical elements. The method we followed is summarized below.

\medskip

\noindent
{\it Oxygen, nitrogen and neon.}
For these elements we have adopted the commonly used schemes:

$$\rm O/H = O^+/H^+ + O^{++}/H^+$$

\noindent
justified by the absence of \heii\lin4686, which is emitted in high-excitation \hii\/ regions and planetary nebulae, where the O$^{3+}$ contribution is non-negligible (\citealt{Kingsburgh:1994});

$$\rm N/O = N^+/O^+$$

$$\rm Ne/O = Ne^{++}/O^{++}$$

\noindent
which derive from the similarity of the ionization potentials of the ions involved (\citealt{Peimbert:1969}).

\bigskip

\noindent
{\it Sulphur and argon.}
We have adopted the metallicity-dependent ionization correction factors (ICFs) of \citet{Izotov:2006}, that 
are based on photoionization model sequences in which the ionizing flux was calculated 
by means of Starburst~99 (\citealt{Leitherer:1999}). The models that are relevant for our work are those labeled
{\it ``high Z"} [\oh\,$>$\,8.2] and {\it ``intermediate Z"} [7.6\,$<$\,\oh\,$\leq$\,8.2] by  \citet{Izotov:2006}, who parameterize the ICFs in terms of the observed \op/O.


\begin{figure}
\medskip
\center \includegraphics[width=0.47\textwidth]{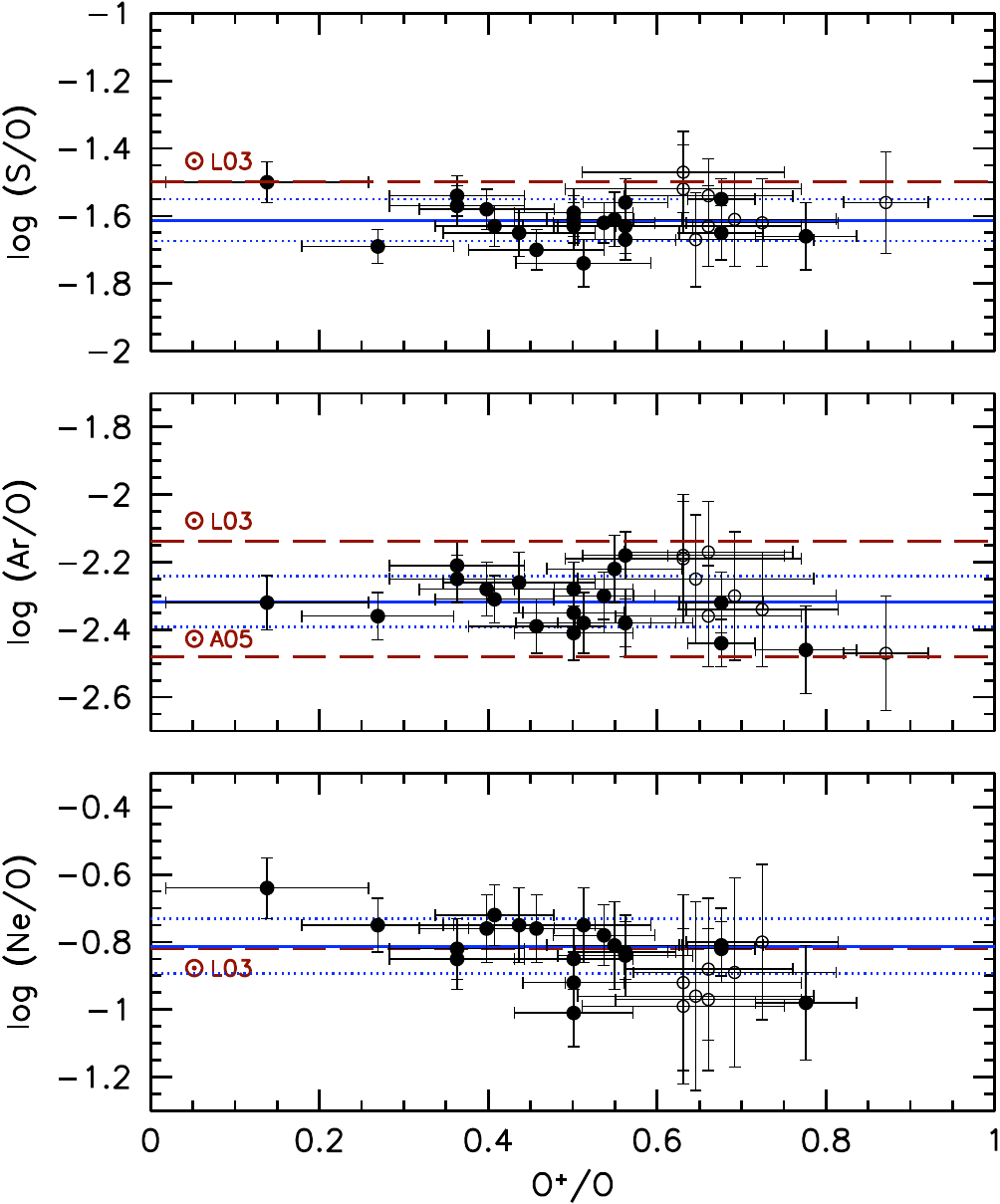}\medskip
\caption{The S/O {\it (top)}, Ar/O {\it (middle)} and Ne/O {\it (bottom)} abundance ratios as a function
of \op/O. The dashed lines represent the solar abundance ratios from \citet[L03]{Lodders:2003} and \citet[A05]{Asplund:2005}. The weighted means calculated from our data points are shown by the continuous horizontal lines, while dotted lines 
are drawn at one standard deviation above and below the mean. Open disk symbols are used for \hii\/ regions 
whose \oiii\lin4363 line is missing, and for which \te\/ was derived from \siii\lin6312 alone. \label{oplus2}}
\epsscale{1}
\end{figure}

\medskip
Fig.~\ref{oplus2} displays, as a function of \op/O, the abundance ratios S/O {\it (top)}, Ar/O {\it (middle)} and Ne/O {\it (bottom)} that are obtained. We use open symbols to represent \hii\/ regions without a \oiii\lin4363 line detection. For these objects the electron temperature of the high-ionization, \opp\/ region was therefore determined
from \siii\lin6312 only, rather than a combination of this line with \oiii\lin4363. In general, the abundance ratios 
derived for these \hii\/ regions have larger error bars, but they agree in our plots with the 
abundance ratios obtained from the use of \oiii\lin4363.
The S/O and Ar/O abundance ratios do not show any appreciable trend with the nebular excitation 
(a linear regression is consistent with zero slope), as expected. Ne/O displays a weak dependence on excitation, which could reflect the inadequacy of the
ionization correction factor that we used. Adopting alternative schemes (e.g.~\citealt{Izotov:2006}; \citealt{Perez-Montero:2007}) results in an steeper dependence than seen in Fig.~\ref{oplus2}. We point out, however, that some
of the Ne/O data points have considerable error bars, and that by removing the single, high-excitation object \#7
the significance of the excitation dependence of Ne/O is reduced considerably. The observational uncertainties and the poor knowledge of the ionization correction for Ne at low excitation prevent us to draw firm conclusions on the behavior of Ne in the \hii\/ regions of NGC~300.

The dashed horizontal lines in Fig.~\ref{oplus2} represent the solar abundance ratios published by \citet[L03]{Lodders:2003}. 
For argon we also display the solar Ar/O ratio taken from \citet[A05]{Asplund:2005}, to show that considerable uncertainty is still present in some of the solar abundance ratios. 
The weighted mean abundances of S and Ne relative to O that we find in NGC~300 (continuous lines) are in good agreement with the solar values by \citet{Lodders:2003}. On the other hand, our mean Ar/O ratio is intermediate between the solar values of \citet{Lodders:2003} and \citet{Asplund:2005}. We note here
that if we had adopted the ionization correction scheme for argon used in earlier works by our group (e.g.~\citealt{Bresolin:2004}), we would have obtained a mean Ar/O value nearly coincident with the \citet{Lodders:2003}
value.
The weighted means and standard deviations that we obtain in our NGC~300 sample are the following:

$$\rm \log (S/O)\,=\,-1.61 \pm 0.06~~~(\odot: -1.50 \pm 0.06)$$
$$\rm \log (Ar/O)\,=\,-2.32 \pm 0.08~~~(\odot: -2.14 \pm 0.09)$$
$$\rm \log (Ne/O)\,=\,-0.82 \pm 0.07~~~(\odot: -0.82 \pm 0.11)$$

\noindent
where the solar values in brackets are taken from \citet{Lodders:2003}. Table~\ref{abundances} summarizes the oxygen
abundances, together with the N/O, S/O, Ar/O and Ne/O abundance ratios.

\medskip
As a consistency check, we have calculated the O/H abundances by using the \oii\lin7325 line, instead of \oii\lin3727, to measure the ionic abundance \op/\hp. This provides a further test of our flux calibration and line flux measurements. We obtained a mean difference log(O/H)$_{7325}$ $-$ log(O/H)$_{3727}$\,=\,
$0.009\pm 0.009$, indicating that no systematic difference exists
between the two measurements. This supports the result obtained by 
\citet{Kniazev:2004} from the spectra of more than 200 \hii\/ galaxies from the Sloan Digital Sky Survey
[log(O/H)$_{7325}$ $-$ log(O/H)$_{3727}$\,=\,$-0.002\pm 0.002$], and the
conclusion that \oii\lin7325 can be used to measure \op/\hp\/ ionic abundances, albeit with larger uncertainties, in those cases where \oii\lin3727 is unavailable (e.g.~\citealt{Izotov:2006}).

\subsection{Trends with O/H}


\begin{figure}
\medskip
\center \includegraphics[width=0.47\textwidth]{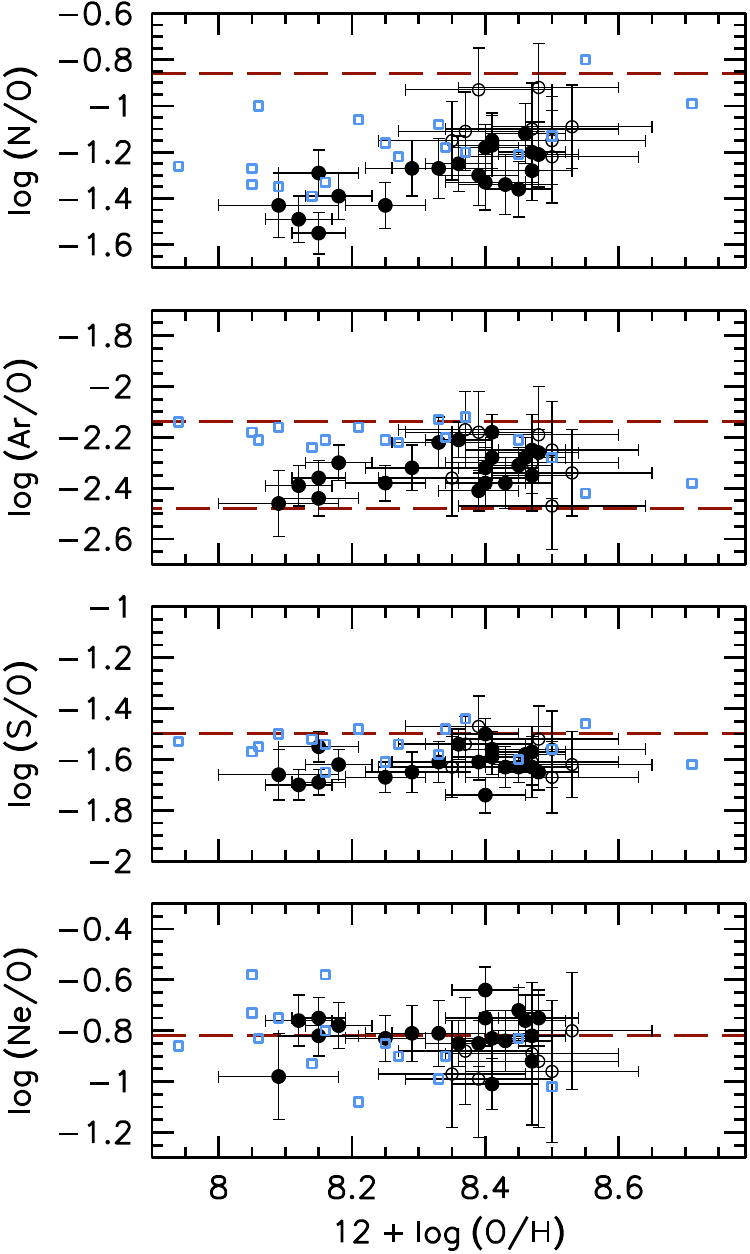}\medskip
\caption{Trends of N/O, Ar/O, S/O and Ne/O with oxygen abundance. In each plot we also include as a comparison sample the M101 data from \citet[squares]{Kennicutt:2003}. The horizontal dashed lines are the solar values shown in Fig.~\ref{oplus2}. Open disk symbols are used for \hii\/ regions whose \oiii\lin4363 line is missing.\label{fig_oh}}
\epsscale{1}
\end{figure}

\begin{figure}
\medskip
\center \includegraphics[width=0.47\textwidth]{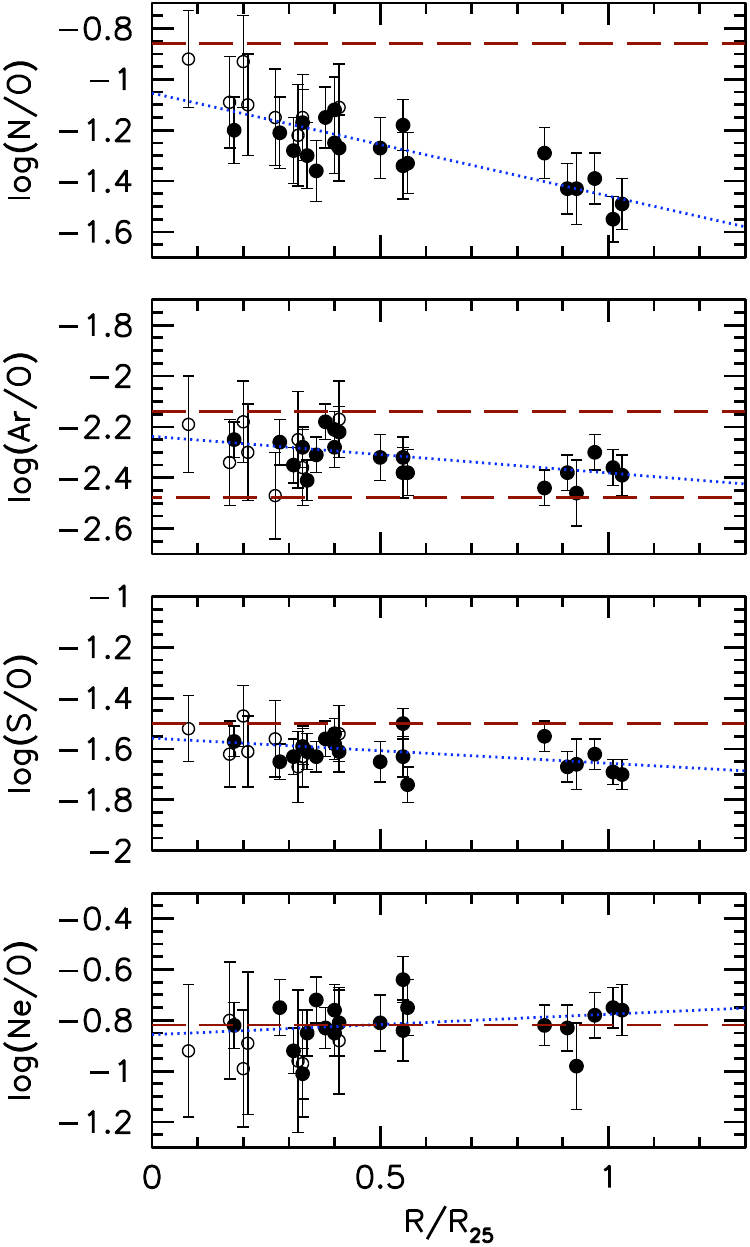}\medskip
\caption{Observed trends of the abundance ratios N/O, Ar/O, S/O and Ne/O with galactocentric distance, in units of the isophotal radius \rtf. The horizontal dashed lines represent the solar values, as in Fig.~\ref{oplus2}. The
dotted lines show the weighted linear regressions to the observations. As in previous figures, open symbols are used for \hii\/ regions whose \oiii\lin4363 line is missing.
\label{fig_r}}
\epsscale{1}
\end{figure}

The relationships between the abundances of N, Ar, S and Ne relative to oxygen ($\log X$/O) and the oxygen abundance \oh\/
are displayed in Fig.~\ref{fig_oh}. The M101 comparison sample from \citet{Kennicutt:2003} is shown with the square symbols. The well-known increase of N/O with O/H, attributed to a secondary component in the nucleosynthesis of nitrogen (\citealt{Vila-Costas:1993}), is apparent in Fig.~\ref{fig_oh}. The independence of the Ne/O ratio on O/H
is also clear from the bottom panel of this figure. Less clear is the situation for Ar/O and S/O. In both cases,
the significance of the correlation with O/H is fairly low. However, if we look at the dependence of the abundance ratios with galactocentric distance, as done in Fig.~\ref{fig_r}, the correlation appears more clearly defined.
The dotted lines represent the weighted linear regressions to the data points. For Ne/O the data are clearly compatible with a constant Ne/O across the disk of the galaxy. The galactocentric dependence of N/O is consistent with the result that this abundance ratio is a function of O/H, as seen in Fig.~\ref{fig_oh}. The Spearman rank correlation coefficient for both S/O and Ar/O as a function of radius is 0.41, and
the slopes are significant at approximately the 3\,$\sigma$ level. 
These trends are opposite in sign to the one detected for S/O in M33 by \citet{Vilchez:1988} and in M51 by \citet{Diaz:1991}. More recent studies on the chemical composition of these two galaxies, however, have failed 
to reproduce these results (\citealt{Bresolin:2004}; \citealt{Magrini:2007}).

Despite the statistical significance of the result, we are reluctant to draw conclusions on the real trend
of S/O and Ar/O with oxygen abundance, with the consequent implication for the nucleosynthetic origin of S and Ar (these elements are thought to be produced by the same massive stars that produce oxygen).
 First of all, we stress that the ionization correction factors for sulphur and argon are still poorly known, and depend strongly on the ionizing properties of the stellar models used to derive them. Despite great progress in stellar atmosphere codes of massive stars in recent years, important differences
in the calculated spectral energy distributions and ionizing output between models still exist (\citealt{Simon-Diaz:2008}). An increase of 15\% in the ICFs at low metallicity [\oh\,$<$\,8.2] would remove the radial trends 
seen for S and Ar in  Fig.~\ref{fig_r}. However, we stress that the result does not depend on the particular choice
of ICFs we made. Adopting the ICF(S) suggested by \citet{Stasinska:1978}, as generalized by \citet{Barker:1980}, and adopted in several recent papers (e.g.~\citealt{Bresolin:2004}; \citealt{Perez-Montero:2006}), or the ICF(Ar) from
\citet{Perez-Montero:2007}, would lead to a similar galactocentric dependence of these two elements.

\citet{Kennicutt:2003} pointed out that the ionization correction scheme they adopted, based on the \citet{Stasinska:1978} formulation, is likely to underestimate the ICF for sulphur at high excitation levels (and their correction for argon is based on the assumption that Ar/S\,=\,\arpp/\spp). However, we do not 
detect a significant radial gradient in excitation, as measured by \op/O. 
We do measure a small gradient in the ionizing stellar temperatures, by means of the radiation softness parameter $\eta$\,=\,(\op/\opp)/(\sulp/\spp) (\citealt{Vilchez:1988a}), in the sense that hotter temperatures are found at larger radii (see \S~\ref{ionizing}), but the quantification of the effect is highly model-dependent.
It is unclear whether this could affect the magnitude of the ionization correction, but we could speculate that 
harder ionizing spectra, as found at large galactocentric distances, could lead to increased proportions of
S$^{3+}$ and Ar$^{3+}$, and therefore larger ICFs.

In alternative, observational uncertainties could be responsible for the observed trends, although we were not able to isolate a mechanism that would systematically decrease S/O and Ar/O with increasing galactocentric distance. One possibility is that the emission line fluxes in the red spectral region are depressed relative to those in the blue. However, as just mentioned, we cannot identify a reason why this effect should depend on galactocentric distance. Moreover, the fluxes of lines in the red are tied to Balmer or Paschen lines, whose intensity follows precise
theoretical (case B) values. 

We close this section by pointing out that, interestingly, a decreasing trend of S/O with radius has been reported in NGC~300 by \citet{Christensen:1997}, although the statistical significance of their result is likely to be quite small, as it rests mostly on a single \hii\/ region at small galactocentric distance.
An S/O ratio that is significantly lower than the average found in the rest of the disk is also observed for an \hii\/ region in M101 located near 
the isophotal radius (\citealt{Garnett:1994}). Unfortunately, sulphur and argon lines were not detected in the outermost ($R$/\rtf\,=\,1.25) \hii\/ region so far spectroscopically observed in M101 (\citealt{Kennicutt:2003}), therefore we cannot confirm that a decrease in S/O exists in this galaxy.


\subsection{Helium}
In order to estimate the He/H abundances in the NGC~300 \hii\/ regions we relied on the measured
intensities of the two  neutral helium lines, \hei\lin5876 and \hei\lin6678. No He$^{++}$ contribution to the total helium abundance
is expected, because high-excitation lines, in particular \heii\lin4686, are not observed in our sample.  We accounted for the absorption line component of the lines arising from 
the underlying stellar population, following the procedure outlined by \citet{Kennicutt:2003}, and which is based on 
the use of the measured equivalent widths of the emission lines. The effect for the NGC~300 sample is 
$<$\,5\% (except for \#22, for which it is 10\%).
The He$^+$/H$^+$ ionic ratios were derived from the line strengths adopting the line emissivities of \citet{Porter:2007} for \hei\/ and \citet{Storey:1995} for \hi, both calculated at the \opp\/ temperature derived from the 
auroral line analysis.

\begin{figure}
\medskip
\center \includegraphics[width=0.47\textwidth]{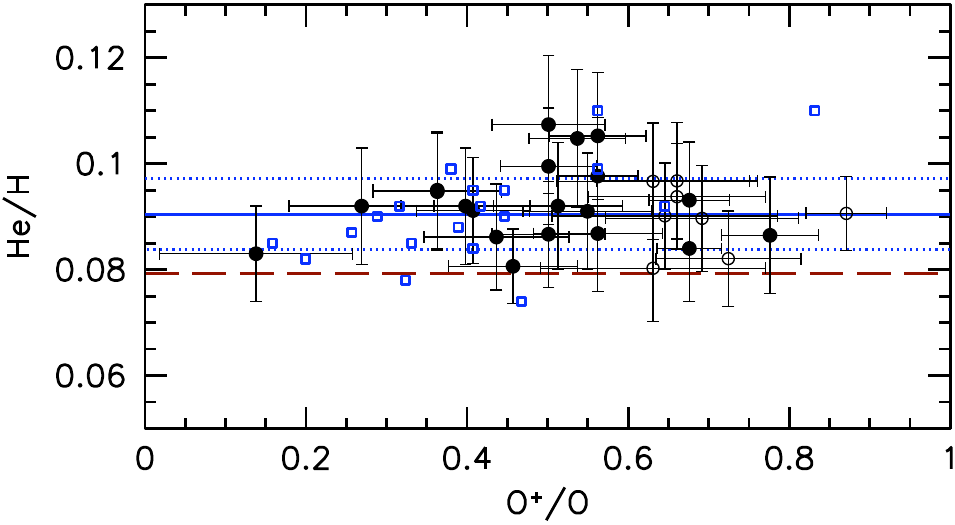}\medskip
\caption{The He/H abundance ratio as a function of excitation \op/O. The horizontal dashed line represents the solar value from \citet{Lodders:2003}. The weighted mean is shown by the continuous horizontal line, with dotted lines 
drawn at one standard deviation above and below the mean. The M101 sample of \citet{Kennicutt:2003} is shown with 
open square symbols.\label{fig_he}}
\epsscale{1}
\end{figure}

\begin{figure}
\medskip
\center \includegraphics[width=0.47\textwidth]{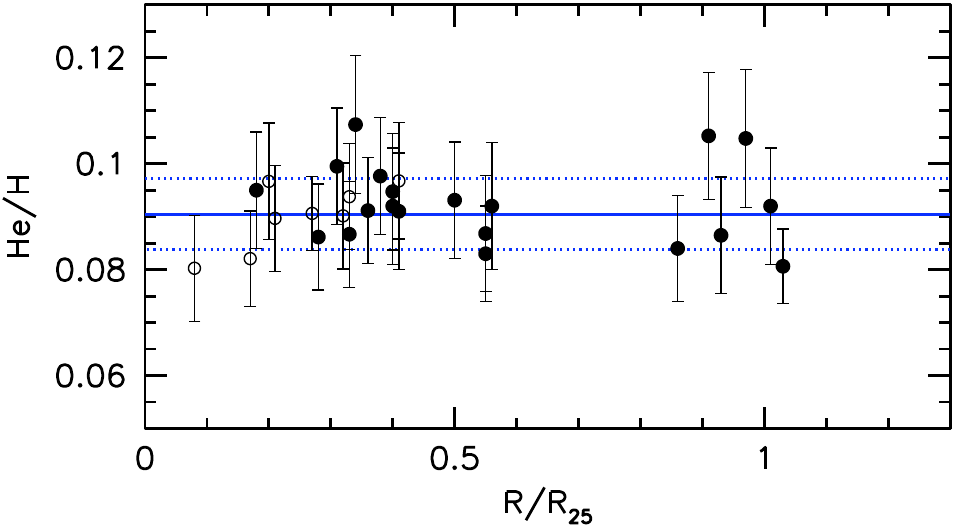}\medskip
\caption{The He/H abundance ratio as a function of galactocentric distance. The weighted mean is shown by the continuous horizontal line, with dotted lines drawn at one standard deviation above and below the mean.\label{fig_he2}}
\epsscale{1}
\end{figure}

The correction for the presence of neutral helium, which can be significant at low excitation, was carried out
using a method based on the radiation softness parameter $\eta$ (\citealt{Vilchez:1989}; \citealt{Izotov:1994}).
We derived an approximate analytical relation between $\eta$ and ICF(He) from the models of \citet{Stasinska:2001}, obtaining

\begin{equation}
\rm ICF(He) = 1.585 + \log\eta\,(1.642\log\eta - 1.948)
\end{equation}

\noindent
valid in the 0.6 $< \log\eta <$\,2.0 interval. For $\log\eta < 0.6$ (harder spectra, or hotter ionizing stars), we 
took ICF(He)\,=\,1. 
Multiplying the He$^+$/H$^+$ ionic ratios by ICF(He) we then obtained the total helium abundances given in Col.~8 of
Table~\ref{abundances} and displayed in Fig.~\ref{fig_he}. The weighted mean He/H\,=\,0.090\,$\pm$\,0.007 is consistent with the observations in our comparison sample in M101 (square symbols). In this galaxy, \citet{Kennicutt:2003} observed an increase of the He/H ratio at the high abundance end, at small galactic radii. We do not see the same effect in NGC~300. Fig.~\ref{fig_he2} shows that the He/H ratio does not correlate with galactocentric distance. The different behavior between the two galaxies could be 
related to the smaller central metallicity in NGC~300 and the modest radial chemical composition gradient in its disk.


\section{The metallicity gradient in NGC~300}

Since the works of \citet{Searle:1971} and \citet{Shields:1974} the radial trends of \hii\/ region chemical abundances in external spiral galaxies have been the subject of many investigations.
Among these we mention the comparative studies, based on strong-line methods, of relatively large samples of galaxies
by \citet{Vila-Costas:1992}, \citet{Zaritsky:1994} and \citet{Pilyugin:2004}, the auroral line-based studies of individual galaxies
by \citet{Garnett:1997} and \citet{Kennicutt:2003}, and the reviews by \citet{Henry:1999} and \citet{Garnett:2004}.
The \hii\/ region abundance gradient in NGC~300 has been measured by several authors, including \citet{Pagel:1979}, \citet{Webster:1983} and \citet{Deharveng:1988}, and has been also re-derived from the data published by those authors by \citet{Vila-Costas:1992}
and \citet{Zaritsky:1994}. However, none of these gradient determinations are based on the detection of
\oiii\lin4363, and were obtained from the \rtwothree\/ indicator.
Following our determination of the direct abundances of 28 \hii\/ regions in \S~4, we present here the metallicity gradient in NGC~300 as obtained from the auroral line method, and compare it with the result from the young stellar content. \\

\noindent{\it \hiiit\/ \it regions~-- }
For the \hii\/ regions we assume, as is customary, that the overall metallicity is
well traced by the oxygen abundance. This is justified by the fact that in \hii\/ regions about half of the atoms in the gas phase heavier than hydrogen and helium are oxygen atoms (e.g.~in the case of the Orion nebula, \citealt{Esteban:2004}).
Carbon can be as abundant as oxygen at high metallicity (\citealt{Garcia-Rojas:2006}), but it is much more difficult to measure in the optical, as it emits only feeble recombination lines. 

An important quantity that affects our comparison with the stellar metallicities is represented by the amount of oxygen that is depleted onto dust grains in ionized nebulae. 
The gas-phase composition of the most common ions of the diffuse ISM in the solar neighborhood is derived from weak UV absorption features measured along sight lines to various stars. For the case of oxygen, observations of \oone~\lin1356 with the Hubble Space Telescope
(\citealt{Meyer:1998,Cartledge:2004}) and other faint \oone\/ transitions with the Far Ultraviolet Spectroscopic Explorer (\citealt{Oliveira:2005}) provide a value for the gas-phase oxygen abundance
near the Sun ($d<1$\,kpc) of \oh\,=\,$8.54\pm0.02$. From the Galactic \hii\/ region radial abundance gradient, 
measured at the solar galactocentric distance, \citet{Pilyugin:2006}  find a good agreement with the
interstellar absorption line results. For the specific case of the Orion nebula, which, at a distance of 389\,pc from the Sun (\citealt{Sandstrom:2007}), can be taken as representative of the gas-phase composition in the solar vicinity, 
\citet{Esteban:2004} obtained \oh\,=\,8.51$\pm0.03$ using the nebular collisionally excited lines, in good agreement with the results above.
However, from {\sc o\,ii} recombination lines the same authors obtained an oxygen abundance higher by 0.14 dex, \oh\,=\,$8.65\pm0.03$, a manifestation of
the abundance discrepancy observed in a number of Galactic and extragalactic \hii\/ regions (\citealt{Garcia-Rojas:2007}). 

Quantifying the amount of oxygen that is locked up into dust grains is made difficult by the uncertainties in the total (gas-phase\,+\,dust) amount that is assumed, and usually taken from the composition of the Sun, B stars or young cool stars (\citealt{Sofia:2004}). The study of the Orion nebula and its surroundings provides
one way of estimating the amount of O depletion, from the comparison of the ionized gas with young B stars. The mean O abundance from the investigation of three B0.5\,V stars in the Trapezium cluster by
\citet{Simon-Diaz:2006} is \oh\,=\,$8.63\pm0.03$. Taking the composition of the diffuse clouds in the solar neighborhood and the collisionally excited line result for the Orion nebula from \citet{Esteban:2004} 
would suggest a depletion factor of approximately $-$0.1 dex. Recently \citet{Przybilla:2008} measured the chemical composition of six early-B stars (luminosity classes {\sc iii-v}) in the solar neighbourhood ($d<500$\,pc), obtaining 
\oh\,=\,$8.76\pm0.03$, which, compared with the diffuse ISM abundance, would imply an oxygen dust depletion of approximately $-0.2$ dex. \citet{Mesa-Delgado:2009} also estimated a $-0.2$ dex depletion factor in the Orion nebula for the
case in which temperature fluctuation in the ionized gas are negleted.

For our comparison in NGC~300, we will adopt $-0.1$ dex as the minimum depletion value that is required to match stellar and gaseous oxygen abundances in the solar vicinity, with the understanding that
it could be as high as $-0.2$ dex, and assuming that the result, obtained for our immediate surroundings in the Milky Way can be generalized to external galaxies.
Furthermore, the caveat remains that the study of recombination lines in \hii\/ regions yield larger abundances (about 0.2 dex) than collisionally excited lines.

\medskip


\begin{figure*}
\medskip
\center \includegraphics[width=1\textwidth]{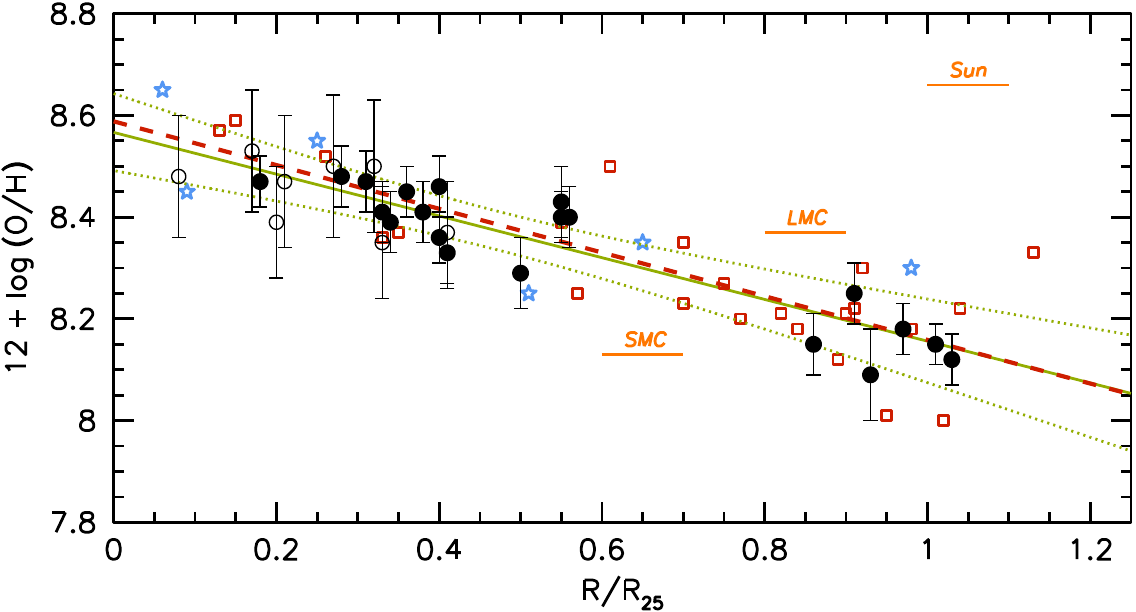}\medskip
\caption{The radial metallicity gradient obtained from \hii\/ regions (circles) and blue supergiants (star symbols: B supergiants; open squares: A supergiants). The weighted regression line to the \hii\/ region data is shown by the
continuous (green) line. The dotted lines show the 95\% confidence level interval for the regression line. The dashed line
represents the weighted regression to the BA supergiant star data. For reference, we include the oxygen abundances of the Magellanic Clouds and the solar photosphere.\label{oradial}\\ \\}
\end{figure*}

\noindent{\it Blue supergiants~-- }
A catalog of nearly 70 early-type supergiant stars in NGC~300 was compiled by \citet{Bresolin:2002a}, based on VLT multi-object spectroscopy, with the goal of providing high-quality stellar spectra for chemical abundance 
follow-up studies. As a proof of concept, these authors analyzed the metallicities of one B9-A0~Ia star and one A0~Ia star.
Later, \citet{Urbaneja:2005} derived abundances of several elements (C, N, O, Mg, Si) in six early-B (B0.5-B3) supergiants, spatially distributed along a wide range of galactocentric distances, allowing them to carry out a first comparison with the oxygen abundance gradient derived from \hii\/ regions. The oxygen gradient derived from the 
six stars was $-0.060 \pm 0.049$ dex\,kpc$^{-1}$ (scaled to our adopted distance to the galaxy), with a central 
abundance of \oh\,=\,8.58\,$\pm$\,0.13. 
However, due to the lack of direct measurements of electron temperatures for the NGC~300 \hii\/ regions, the nebular abundances were estimated from the application of several calibrations of the \rtwothree\/ strong-line abundance indicator to the emission line fluxes compiled by \citet{Deharveng:1988}. Significantly different nebular abundance gradients were obtained by varying 
the adopted \rtwothree\/ calibration, both in terms of slope and zero-point, making it difficult to reach meaningful conclusions regarding the comparison between the stellar and nebular galactocentric abundance trends.

More recently, \citet{Kudritzki:2008} analyzed a sample of 24 A-type supergiants (spectral types from B8 to A4) drawn from the \citet{Bresolin:2002a} catalog, deriving stellar parameters and metallicities using a grid of line-blanketed stellar models and NLTE line formation calculations. While in this case the authors did not derive abundances of individual elements, because at the low resolution of the VLT spectra the features in A supergiant spectra are mostly blends of iron, titanium, chromium and other lines, the fitting of features across the whole wavelength range covered was found to be quite sensitive to the choice of the model metallicities (the typical estimated uncertainty was 0.2~dex).
In the construction of the models the solar abundance pattern of the various chemical elements was assumed, from \citet{Grevesse:1998}. The metallicities of the A supergiants obtained by \citet{Kudritzki:2008} refer therefore 
to the abundances of a variety of heavy elements (Mg, Si, S, Ti, Cr, Fe) combined, whose individual abundance ratios reflect the solar abundance pattern. While this restriction can be relaxed in future modeling, allowing, for example, the investigation of variations of the $\alpha$/Fe element ratio in galaxies, the metallicity values obtained in the case of the NGC~300 A supergiants should be regarded as representative of the overall metal content in the stellar atmospheres of these young stars. 

\bigskip
In comparing the metallicities from the supergiants stars with  the oxygen abundances from the \hii\/ regions 
we make the assumption that for the A stars oxygen scales with metallicity, and that the solar metallicity value corresponds to \oh$_\odot$\,=\,8.66 (\citealt{Asplund:2005}). Fig.~\ref{oradial} shows the radial metallicity gradient determined from the \hii\/ regions (circles) and blue supergiants (star symbols: B supergiants; open squares: A supergiants; galactocentric distances for the stars have been recomputed adopting the galaxy parameters
in Table~\ref{parameters}). For clarity the error bars for the stellar data are omitted, but typical uncertainties
are on the order of 0.2 dex.
As a reference, we include in the plot the mean level of the nebular oxygen abundances measured in the Magellanic Clouds by \citet{Russell:1990}, as well as the solar value from  \citet{Asplund:2005}. In the figure we have omitted the outlier A supergiant A10 from \citet{Kudritzki:2008}. 
As in previous figures, we use open circles for \hii\/ regions that have no \oiii\lin4363 detection, and therefore the electron temperature in their O$^{++}$-emitting region was derived from \siii\lin6312. The corresponding oxygen abundances have larger errors than those for the \hii\/ regions for which we have used both 
\oiii\lin4363 and \siii\lin6312 to obtain \te. 

A weighted linear regression to the \hii\/ region data, with weights equal to the reciprocal of the variance, yields:
\begin{equation}
{\rm \oh_{gas} = 8.57~(\pm 0.02) - 0.41~(\pm 0.03)}~R/R_{25} \label{regression}
\end{equation}
\noindent
and is shown with a continuous line in Fig.~\ref{oradial}. 
The weighted regression to the combined A and B supergiant data, using the variances from \citet{Kudritzki:2008} and \citet{Urbaneja:2005}, is:
\begin{equation}
{\rm \oh_{stars} = 8.59~(\pm 0.05) - 0.43~(\pm 0.06)}~R/R_{25}
\end{equation}
\noindent
and is shown with a dashed line in Fig.~\ref{oradial}. A straightforward error analysis shows that the slopes and the intercepts of the two regressions are not significantly different.
The virtual coincidence of the slopes of the nebular and stellar abundance gradients is remarkable.
This result is quite robust, since it does not depend on the somewhat uncertain dust depletion factor, which affects the absolute abundance values, and the possible effects of temperature fluctuations on the derived abundances, because they appear to be independent of metallicity.
From the \hii\/ region regression, which has the smaller errors, we derive an oxygen abundance gradient
of $-0.077\pm0.006$ dex\,kpc$^{-1}$, while the blue supergiants yield $-0.081\pm0.011$ dex\,kpc$^{-1}$ (the small difference relative to the value published by \citealt{Kudritzki:2008}, $-0.083$ dex\,kpc$^{-1}$, is due to the slightly different orientation parameters used to calculate the galactocentric distances).

The intercepts of the linear fits are virtually coincident.
Due to our initial assumption that the A supergiant oxygen abundances  scale with the metallicity obtained from
the analysis of the line blends in their spectra,   the vertical positions of the A supergiant  
data in Fig.~\ref{oradial} depend on the choice of the solar oxygen abundance, whereas the \hii\/ region abundances
do not. However, in Fig.~\ref{oradial} we do not detect a significant offset between the A supergiants and the B supergiants, which provide direct measurements (albeit model-dependent) of the oxygen abundance via the stellar {\sc o\,ii} spectral features. In any case, it is important to remember that additional 
effects, such as the uncertainties of the atomic parameters on the derived abundance, also influence the comparison of the absolute values of the chemical abundances. For example, the default atomic data used by 
{\sc iraf}'s {\it nebular} tasks distributed with {\sc stsdas} Version 3.8 (Feb 2008), which had not been updated since 1997, would yield a mean O/H value lower by about 0.04 dex than reported here.

The agreement we find between nebular and stellar abundances leaves little room for effects that would  systematically increase the nebular abundances, such as metal depletion onto dust grains, or large-scale temperature fluctuations. By considering the modest dust depletion factor of $-0.1$ dex for oxygen discussed earlier the intercept of the linear regression to the \hii\/ region data for O/H would still be consistent 
with the A supergiant fit, within the 1$\sigma$ uncertainties. 
We estimate that the stellar and nebular intercepts would differ at the 95\% confidence level once we reach
\oh$_{\rm gas}$\,=\,8.70 (neglecting other sources of systematic errrors, such as 
uncertainties in the atomic parameters and the solar metallicity).
This would leave no room for temperature fluctuation effects.
Clearly, measurements of metal recombination lines in this galaxy would be helpful to secure constraints on the abundance discrepancy factor in NGC~300.


We do not see any evidence in Fig.~\ref{oradial} for a break in the abundance gradient of NGC~300, in particular for a steepening of the gradient in the inner disk, which was suggested by \citet{Vila-Costas:1992},
who applied the \rtwothree\/ index on a compilation of published data to derive chemical abundances.
This result was not confirmed by \citet{Zaritsky:1994}, who could not exclude that some of the breaks in the line ratios observed for a few galaxies  at certain galactocentric distances might not represent actual bends in the abundance gradients (as more recently concluded by \citealt{Pilyugin:2003c}).
A central steepening might be present in other galaxies, for example M33 (\citealt{Vilchez:1988,Magrini:2007}), but a simple exponential function appears to be sufficient to describe the dependence of O/H on radius in NGC~300. 
\citet{Zaritsky:1994} defined the `characteristic' oxygen abundance of a galaxy as the value measured at 0.4\,\rtf. This quantity correlates with the integrated galaxy metallicity 
(\citealt{Moustakas:2006}).
From Eq.~\ref{regression} we find a characteristic abundance \oh\,=\,$8.41\pm0.04$ for NGC~300.

The distribution of the \hii\/ region data points in Fig.~\ref{oradial} has quite a small dispersion, with an rms residual from the linear fit of only 0.05 dex, which is similar to the measurement uncertainty of the most accurate O/H abundances presented in Table~\ref{abundances}
(the reduced $\chi^2$ statistics equals 1.2, implying that the abundance error estimates are realistic).
The scatter is smaller than what has been observed from auroral line measurements in other galaxies, such as M101 (\citealt[0.09 dex]{Kennicutt:2003}, comparable to the measurement errors), and especially M33, where \citet{Rosolowsky:2008} found a dispersion which is considerably larger than the measurement errors, with an intrinsic variance of 0.11 dex (from their data, the rms scatter computed in the same way as in the case of NGC~300 is 0.16 dex). In NGC~300 the observed oxygen abundance scatter at a given radius is therefore consistent with the measurement errors. This suggests that azimuthal abundance variations are negligible at the level of precision
attained in this work. However, the number of \hii\ regions observed should be increased considerably, especially in the outer disk, in order to test for azimuthal variations.

\section{Discussion}

\subsection{Comparisons between gas and stars in other galaxies}
In this section we briefly summarize the main results regarding the comparison of chemical abundances
obtained from \hii\/ regions and blue supergiants in nearby galaxies, in order to put our result for NGC~300 into
a wider context.
We limit our comparisons to the abundances of oxygen, since this is, in most cases, the only chemical element that can be reliably measured in both gas and stars (nitrogen is subject to evolutionary effects, i.e.~enrichments and depletions, in early-type stars). We also do not strive for an exhaustive review of the subject, and concentrate only on \hii\/ regions as representative of the gaseous abundances, since in NGC~300 planetary nebulae abundances have not been published yet (but a companion paper, by Pe{\~n}a et al.~2009, is in preparation).

Some of the previous comparisons in external galaxies has suffered
from the lack of direct abundances (those based on \te\/ determinations ) of \hii\/ regions, and had therefore to deal 
with the systematic uncertainties introduced by the use of different strong-line abundance indicators. This was the case in the previous study of six B supergiants in NGC~300 by \citet{Urbaneja:2005}, or in the analysis of B, A and F supergiants in M31 (\citealt{Smartt:2001}; \citealt{Trundle:2002}; \citealt{Venn:2000}). Firmer results
have been obtained in nearby dwarf galaxies, in which the low oxygen abundances favor, due to the higher electron temperature resulting from smaller gas cooling, the detection of \te-sensitive auroral lines, in particular \oiii\lin4363. In most cases, a very good agreement is found between the oxygen abundance from B-type stars and \hii\/ regions (Magellanic Clouds: \citealt{Trundle:2005, Hunter:2007} - WLM and IC~1613: \citealt{Bresolin:2006a, Bresolin:2007a,Urbaneja:2008} - NGC~6822: \citealt{Lee:2006} - NGC3109: \citealt{Evans:2007,Pena:2007}). However, the metallicity
in these small, chemically homogeneous systems is low, typically below  the SMC value \oh\,=\,8.1 (1/4 solar).

The comparison at higher metallicities becomes more challenging, as explained above, due to the difficulty of measuring reliable \hii\/ region chemical abundances as the nebular cooling efficiency increases (by contrast, stellar features become more pronounced). Metallicities that are larger than those encountered in dwarf galaxies can be found in the central regions of spirals (obviously we are not considering ellipticals), but the number of target galaxies that are suitable for a comparison with the stellar studies is quite limited. In fact, prior to our work in NGC~300, comparisons have been carried out only in the Milky Way and in M33 (todate no direct nebular abundances are available in M31).
In both cases the picture is not as clear as one would hope. In the Milky Way, where both \hii\/ regions and B stars can be traced in the optical preferentially at galactocentric distances larger than about 6~kpc, due to dust obscuration in the direction of the Galactic center, B-type main sequence stars
have been found to define a radial oxygen abundance gradient of $-0.067\pm0.008$ dex\,kpc$^{-1}$ (\citealt{Rolleston:2000}; see also \citealt{Smartt:1997} and \citealt{Gummersbach:1998}) that well matches the gradient obtained for \hii\/ regions by
various authors, including \citet{Shaver:1983} and \citet{Afflerbach:1997}. On the other hand, more recent studies of \hii\/ regions (\citealt{Deharveng:2000}) and B stars (\citealt{Daflon:2004}) find significantly flatter slopes for the oxygen gradient, $-0.039\pm0.005$ and $-0.031\pm0.012$ dex\,kpc$^{-1}$, respectively, i.e.~about a factor of two smaller. 
These works underline the need for accurate nebular electron temperatures and for a self-consistent NLTE analysis of homogeneous samples of OB stars, in order to improve chemical abundance determinations in these young populations. Comparisons in single, nearby star-forming regions appear particularly important
in assessing the reliability of the analysis methods. A recent example is given by \citet{Simon-Diaz:2006} who, using the latest generation of hot star models,  have found 
a remarkable agreement in the oxygen surface content of three Trapezium cluster B main-sequence stars
[average \oh\,=\,8.63\,$\pm$\,0.03] and of the gas-phase abundance of the Orion nebula 
[\oh\,=\,8.65\,$\pm$\,0.03], analyzed from high-resolution spectroscopy by \citet{Esteban:2004}.

Comparative studies of \hii\/ region and hot star abundances in M33 started with
\citet{McCarthy:1995} and \citet{Monteverde:1996}, who established with the first metallicity determinations for A and B supergiants in this galaxy a rough agreement with the nebular abundances. The more recent quantitative B supergiant 
work by \citet{Urbaneja:2005a} derived a linear radial oxygen abundance gradient of $-0.06\pm0.02$ dex\,kpc$^{-1}$, which is in agreement with the \hii\/ region O/H gradient of $-0.054\pm0.011$ dex\,kpc$^{-1}$ measured across the whole optical disk of the galaxy by \citet{Magrini:2007}. The latter authors, following \citet{Vilchez:1988} and \citet{Urbaneja:2005a}, proposed that the combined nebular+stellar gradient could be better represented by a steeper portion in the inner ($R<3$\,kpc) disk relative to the outer part of the galaxy.
However, the picture is complicated by additional \hii\/ region studies, reporting slopes of the O/H gradient that are both steeper ($-0.12\pm0.02$ dex\,kpc$^{-1}$: \citealt{Vilchez:1988}) and shallower 
($-0.012\pm0.011$ dex\,kpc$^{-1}$: \citealt{Crockett:2006}; $-0.027\pm0.011$ dex\,kpc$^{-1}$: \citealt{Rosolowsky:2008}). These last authors suggest that the different determinations can, in fact, be compatible with each other, because the errors in the slopes are likely to be under-estimated in the case of a large intrinsic scatter of the abundances around the mean gradient. We add that the blue supergiants
studied by \citet{Urbaneja:2005a} provide $\sim$0.3 dex systematically larger abundances than the \hii\/ regions studied by \citet{Magrini:2007}.

In conclusion, there is a still rather small body of evidence that in general metallicity determinations 
carried out by means of \hii\/ regions and hot stars (B dwarfs and supergiants, and A supergiants) in galaxies
are in rough agreement, but some apparently significant differences are borne out when looking at the details.
The specific cases of the abundance gradient in the two best-studied spiral galaxies, the Milky Way and M33,  reveal unacceptable inconsistencies between \hii\/ region abundances from different authors, that are hard to explain simply in terms of measurement errors. A critical aspect of the nebular work is represented by the accuracy of the electron temperatures, which are usually derived from optical auroral lines (hydrogen radio recombination lines can also be used in the Milky Way, \citealt{Shaver:1983}). 
This problem can be partially circumvented by deriving abundances from mid-IR (e.g.~\neii\lin12.8\,$\mu$m, \neiii\lin15.6\,$\mu$m, \siii\lin18.7\,$\mu$m, \siv\lin10.5\,$\mu$m) and far-IR lines (e.g.~\oiii\llin52,88 $\mu$m),
which are insensitive to \te. This method has been pursued with success in the Milky Way (\citealt{Martin-Hernandez:2002,Rudolph:2006}), but has been rarely applied to derive abundance gradients for external spiral galaxies. A recent example is the study of S and Ne abundances in M33 by \citet{Rubin:2008}, carried out with the Spitzer Space Telescope (unfortunately, the oxygen lines lie outside of the spectroscopic range covered by Spitzer).


\begin{figure}
\medskip
\center \includegraphics[width=0.47\textwidth]{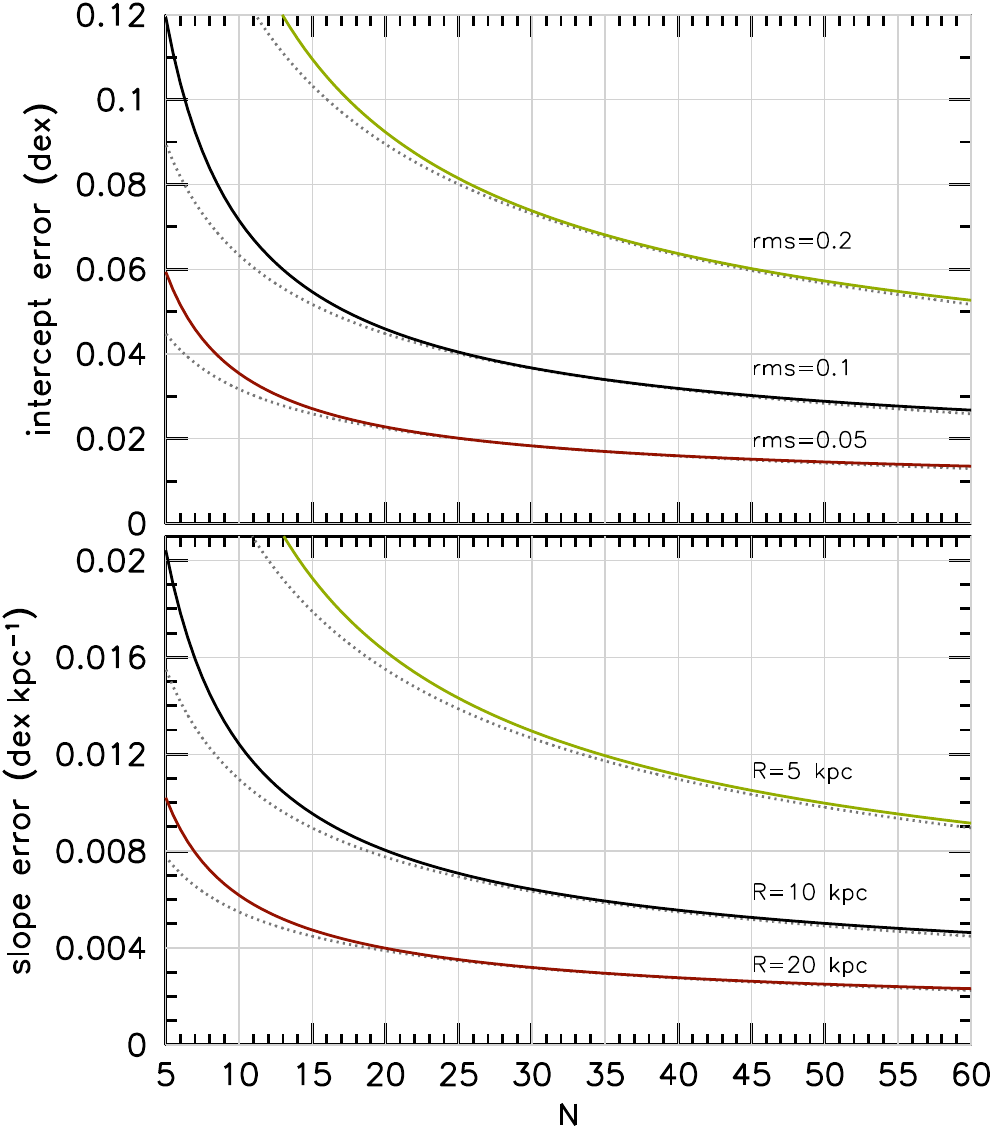}\medskip
\caption{Results of Monte Carlo simulations of the errors measured in the intercept {\it (top)} and the slope {\it (bottom)} of exponential fits to the abundance distribution of \hii\/ regions uniformly distributed in radius within a spiral galaxy as a function of the number of data points. Curves for the intercept error 
are drawn for three different values of the rms scatter (0.05, 0.1 and 0.2 dex). The curves in the bottom panel are drawn for a single rms scatter of 0.1 dex, and three different sizes of the galaxy (5, 10 and 20 kpc).
The analytical results of \citet{Dutil:2001} are shown by the dotted lines.
 \label{simul} \\ \\}
\end{figure}

Sample size can also be an issue in deriving reliable \hii\/ region abundance gradients, especially if the intrinsic abundance scatter is large (as suggested by \citealt{Rosolowsky:2008} for the case of M33).
\citet{Dutil:2001} estimated that a minimum of 16 \hii\/ regions is required in order to obtain zeropoints with an accuracy of $\pm0.05$ dex or better when the rms scatter is 0.1 dex. We have run a series of Monte Carlo simulations to independently estimate the errors in both the intercept and the slope of a single exponential fit to data points distributed uniformly in galactocentric radii, for different values of the scatter and the size of the galaxy. We essentially confirm the results of \citet{Dutil:2001}, except that as the number $N$ of objects considered becomes smaller our simulations predict progressively larger errors (see Fig.~\ref{simul}). The top panel of Fig.~\ref{simul} shows that for an intrinsic rms abundance scatter of 0.05 dex (as appears to be the case of NGC~300, as shown here, and the Milky Way, \citealt{Deharveng:2000}) 10 \hii\/ regions
are sufficient to determine the intercept of the fit to better than 0.04 dex, but the required $N$ rises quickly, with the square of the intrinsic rms scatter.
The curves in the bottom panel for the slope error as a function of $N$ were drawn for a constant rms scatter of 0.1 dex, and different galactic radii (5, 10 and 20~kpc). The slope error at a given $N$ is the same
for a given ratio between these two quantities. Measuring at the 3$\sigma$ level a shallow abundance gradient (0.02-0.03 dex\,kpc$^{-1}$) could require more than 30 \hii\/ regions, as in the case of a galaxy with a radius of 10~kpc and a scatter of 0.1 dex (black curve).


\begin{figure}
\medskip
\center \includegraphics[width=0.46\textwidth]{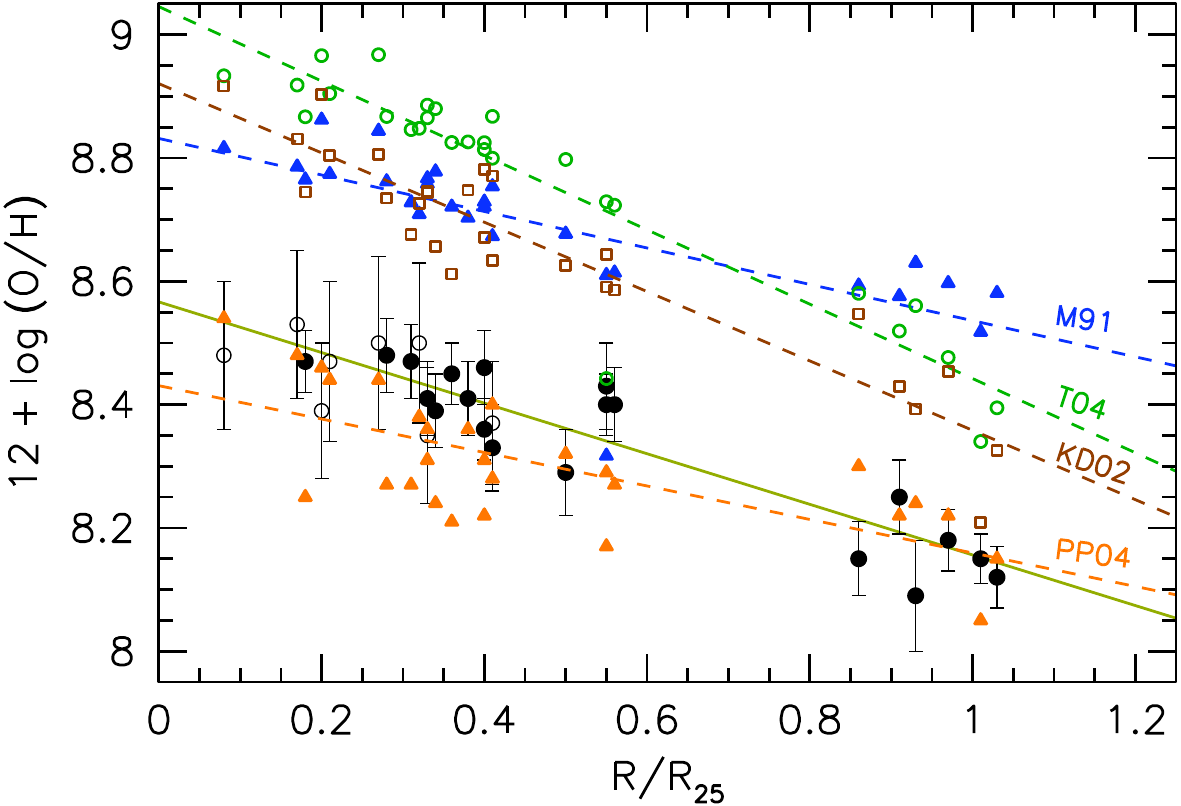}\medskip
\caption{Galactocentric distribution of the abundance values obtained from different strong line methods and calibrations: \rtwothree\/ (\citealt{McGaugh:1991}\,=\,M91, blue triangles; \citealt{Tremonti:2004}\,=\,T04, green circles), \nii/\oii\/ (\citealt{Kewley:2002}\,=\,KD02, open squares), and N2 (\citealt{Pettini:2004}\,=\,PP04, orange triangles). Linear least squares fits  are shown by the dashed lines, and labeled with the appropriate reference. The direct abundances determined 
from our work are shown by the full and open circle symbols, and the corresponding linear fit is shown by the continuous line (same as in Fig.~\ref{oradial}).\label{strongline}}
\end{figure}

\subsection{Strong-line abundances}\label{strong}
Comparisons of  strong-line abundance
determination methods for \hii\/ regions have already been carried out by several authors, and we refer the reader to the raher vast recent literature on the subject (\citealt{Kennicutt:2003}; \citealt{Perez-Montero:2005}; \citealt{Nagao:2006}; \citealt{Yin:2007}; \citealt{Liang:2007a}; \citealt{Shi:2007}; \citealt{Kewley:2008}). In the specific case
of NGC~300 \citet{Urbaneja:2005} have already shown how the adoption of three popular 
calibrations of the \rtwothree\/ method affects the nebular abundances in relation to the B supergiant oxygen abundances. We can also look at the abundance gradients determined by various authors. As an example,
using their \rtwothree\/ calibration, \citet{Zaritsky:1994} obtained a slope of $-0.61\pm0.05$\,\rtf$^{-1}$, about 50\% larger than our value, and an extrapolated central abundance  \oh$\rm_o$\,=\,8.97, compared to our value of 8.57. \citet{Deharveng:1988}, with the \citet{Dopita:1986} \rtwothree\/ calibration, obtained virtually the same values (slope: $-0.63$, intercept: 8.95). \citet{Vila-Costas:1992}, with yet another \rtwothree\/ calibration, otained a slope of $-0.49$\,\rtf$^{-1}$ and an intercept \oh$\rm_o$\,=\,8.78 (using their one-component fit rather than the two-part gradient).
As made clear by a few empirical studies (\citealt{Kennicutt:2003,Bresolin:2004}), the direct abundances obtained from the \te-sensitive auroral lines provide values in the metal-rich, upper branch of \rtwothree\/ that are considerably lower than previous \rtwothree\/ calibrations. The effect appears to be more significant 
at high metallicity, so that the slopes derived from the direct method are shallower. In fact, the gradient derived by \citet{Pilyugin:2004} from the published data using the P-method, which is empirically tied to \oiii\lin 4363 abundances in other galaxies, is in good agreement with our direct measurements (slope: $-0.40$, intercept: 8.49).

We can now illustrate how different strong-line methods compare to the \te-based direct method, as applied to the \hii\/ regions in NGC~300. For simplicity, we limit our choice to only three methods: 
{\it (i)} \rtwothree\,=\,(\oii\lin3727 + \oiii\llin4959,5007)/H$\beta$, with the popular theoretical calibrations by \citet[using the analytical forms
provided by \citealt{Kuzio-de-Naray:2004}]{McGaugh:1991}, and \citet[the analytical fit in their Eq.~1]{Tremonti:2004}, {\it (ii)} the theoretical prediction for the \nii\lin6583/\oii\lin3727 indicator by \citet{Kewley:2002}, and {\it (iii)} N2\,=\,log(\nii\lin6583/H$\alpha$), empirically calibrated by \citet{Pettini:2004}. The output abundances of these methods and their relative calibrations span the abundance range typically covered by strong-line indicators. Fig.~\ref{strongline} shows the abundances that result
from the application of these different indicators and calibrations, together with the direct abundances derived in \S~4, as a function of galactocentric distance of the \hii\/ regions. For each method we include the linear least-square fit to the data. 

Fig.~\ref{strongline} illustrates the well-known result that calibrations of strong-line indices based on theoretical modeling can provide abundance estimates that are larger, by several tenths of a dex,
than those from empirical calibrations (\citealt{Bresolin:2004}). The better agreement between our direct abundance gradient and the gradient obtained from N2, calibrated by \citet{Pettini:2004}, is no surprise, since this calibrations is directly tied to \oiii\lin4363 measurements in extragalactic \hii\/ regions. For \rtwothree\/ we have assumed that all \hii\/ regions, including those at large galactocentric distance, belong to the upper branch of the calibration, even if their \nii/H$\alpha$, the usual discriminator between lower and upper branch, is around the turnover value (around \nii/H$\alpha$\,=\,$-1.1$ to $-1.3$, \citealt{Kewley:2008}). Adopting the lower branch calibration would move the outer data points down by $\sim$0.4 dex, bringing them close to the empirical results, but producing a much steeper 
gradient or an abundance jump from the inner half of the disk. A plot of O/H \vs~\rtwothree\/ (not shown)
suggests that the assumption that all objects belong to the upper branch is justified.

The \hii\/ region \#7 (=\,De\,30) in our list (Table~1) is a clear outlier in relation to the abundance
gradient measured with the \rtwothree\/ method, as it lies 0.35~dex below the corresponding regression line.
This object is characterized by a very hard spectrum, with \oiii/\oii\,=\,9.1, that makes it stand out
also in the traditional BPT (\citealt{Baldwin:1981}) diagnostic diagrams. This is likely related 
to its WR star content (see \S~\ref{wrsect}). The measured log\,\rtwothree\,=\,1.13 is larger than the maximum value considered by the \citet{McGaugh:1991} models, and this is the likely explanation for the failure of the analytical expression used to derive O/H from \rtwothree. This object's oxygen abundance does not appear to be peculiar when measured from the high S/N detection of \oiii\lin4363, or from either
N2 or \nii/\oii.

Fig.~\ref{strongline} also illustrates the fact that the slope of the radial abundance gradient can
be significantly different from the one estimated from the electron temperature method. For example, the slope determined from \nii/\oii\/ is $-0.56\pm0.04$ dex\,\rtf$^{-1}$, compared to $-0.41\pm0.03$ dex\,\rtf$^{-1}$
from the direct method. 
In conclusion, we confirm ealier findings that the direct method yields oxygen abundances that are 
considerably smaller than some popular calibrations of
strong-line methods obtained via grids of theoretical models. While the origin of the discrepancy remains unclear, the direct method result is rather robust, and is backed by the study
of blue supergiants. This has obvious consequences for the study of metallicity gradients in galaxies, as exemplified in Fig.~\ref{strongline}, as well as for the derivation of chemical abundances in star-forming galaxies at high redshift.

\subsection{Wolf-Rayet stars}\label{wrsect}
The detection of Wolf-Rayet (W-R) stars in extragalactic \hii\/ regions is
quite common. The broad lines emitted in the extended atmospheres of these hot, evolved massive stars are easily
discerned even at relatively low S/N ratios in the spectra of the ionized nebulae that harbor them.
The first detection of W-R features in NGC~300 was carried out by \citet{Dodorico:1983}, who found the characteristic 
'blue bump' around 4650~\AA\/ in two \hii\/ regions (our \#5 and \#19). Additional W-R stars have been reported by \citet{Deharveng:1988}, \citet{Schild:1991,Schild:1992}, \citet{Breysacher:1997},
\citet{Bresolin:2002b} and \citet{Schild:2003}, whose census includes 60 known or 
candidate W-R stars, 21 of which have spectroscopic confirmation. As \citet{Schild:2003} pointed out, 
12 of the spectroscopically confirmed W-R stars are of type WC, and they estimated that the overall N(WC)/N(WN) number ratio is $\geqslant 1/3$. This puts NGC~300 approximately in the expected location in the empirical N(WC)/N(WN) \vs~O/H diagram, even though, as a consequence of our improved results for the metallicity in this galaxy, the  abundance value adopted for the central region by \citet{Crowther:2007} should be revised downwards by approximately 0.1--0.15~dex to \oh\,=\,8.5 [the value adopoted by \citealt{Schild:2003} was instead \oh\,=\,8.8].

\begin{deluxetable}{cccl}
\tabletypesize{\scriptsize}
\tabletypesize{\footnotesize}
\tablecolumns{4}
\tablecaption{W-R star detections\label{wr}}

\tablehead{
\colhead{\phantom{aaaa}ID\phantom{aaaa}}	     &
\colhead{\phantom{a}Type\phantom{a}}       &
\colhead{\phantom{a}Reference\phantom{a}}       &
\colhead{\phantom{a}Comment\phantom{a}}	 \\[1mm]
\colhead{(1)}	&
\colhead{(2)}	&
\colhead{(3)}	&
\colhead{(4)}	}
\startdata
\\[-2mm]
5\dotfill  & WC	 &   D\,2     &  \\       
7\dotfill  & WNE	 & ST\,8  &  \\       
12\dotfill  & WNL	 &   & new detection \\       
14\dotfill  & WC4, WN	 & S\,2-4  &  in nearby regions De\,53B-C\\       
17\dotfill  &	 & S\,7-8  &   \\       
19\dotfill  & WC+WN	 & S\,11-12 &  \\       
20\dotfill  &	 & S\,18  & \\       
23\dotfill  &	 & S\,35-37 &   \\       
24\dotfill  & 	 & S\,36  &  \\       
26\dotfill  & WN7	 & S\,49-50-51  &        
\enddata
\tablecomments{Col.~(1): \hii\/ region identification (from Table~\ref{sample}). Col.~(2): 
W-R classification (when available). Col.~(3): source of W-R data. S: \citet[only this reference is provided when
available]{Schild:2003}; D: \citet{Dodorico:1983}; ST: \citet{Schild:1992}. Col.~(4): De\,=\,\citet{Deharveng:1988}.\\}
\end{deluxetable}

We have detected W-R star features in 10 \hii\/ regions of our sample, mostly the 4650~\AA\/ bump (composed 
of \heii\lin4686 and \niii\lin4634--41 and/or \nv\lin4603--20), but 
also \civ\lin5808 (\# 5\,=\,De\,24) and \ciii\lin5996 (\#19\,=\,De\,77). Our spectrum of \#7 (=\,De\,30, already briefly discussed in \S~\ref{strong}) 
also contains broad \heii\llin5200, 4541 and 5411 from the embedded W-R star. This object is absent from the list of 
\citet{Schild:2003}, but corresponds to star 8 of \citet{Schild:1992}, who classified it as an early WN.
We confirm this classification, based on the strong \heii\lin4686, and the presence of \nv\lin4603--20.
For the \heii\lin4686 line of this star we measure a luminosity of $3.8\times10^{35}$ erg\,s$^{-1}$ (for $D$\,=\,1.88~Mpc), 
an equivalent width of 175~\AA\/ and a {\sc fwhm} of 35~\AA. With these measurements we find very good agreement 
with the properties of WN3--4 stars in the LMC (of comparable metallicity) presented by \citet{Crowther:2006b}.

We summarize in Table~\ref{wr} the W-R star detections in our \hii\/ region sample, providing spectral types and references from previous work. We report a new detection in \#12 (=\,De\,45), and assign a WNL classification, based on the presence of \niii\lin4634--41, of comparable strength to
\heii\lin4686, and the absence of \nv\lin4603--20 in emission. 
We also uncovered the blue bump feature in \#14, which corresponds to De\,53A in \citet{Deharveng:1988}, while
\citet{Schild:2003} report W-R stars in the nearby regions De\,53B and De\,53C.



\begin{figure}
\medskip
\center \includegraphics[width=0.47\textwidth]{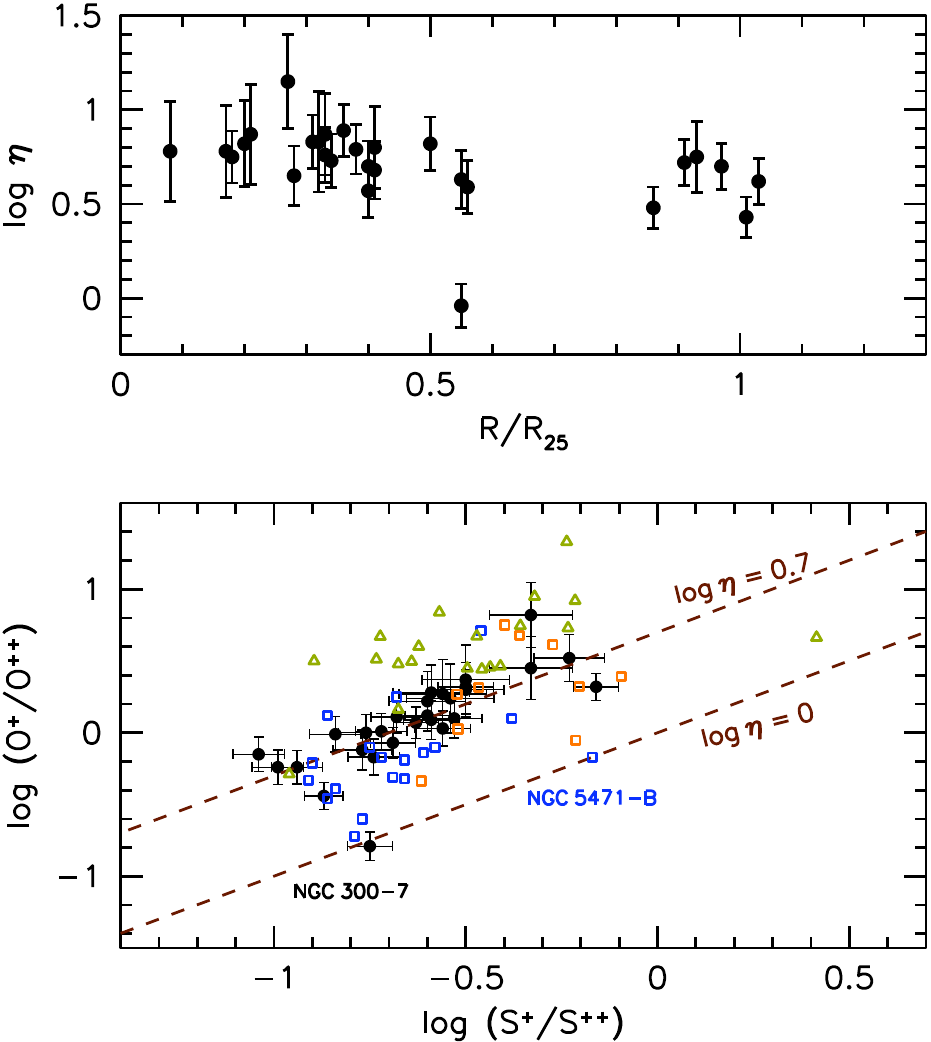}\medskip
\caption{{\it (Top)} Radial gradient of log\,$\eta$\,=\,log\,(\op/\opp)/(\sulp/\spp) in NGC~300. The outlier at
$\log\eta$\,$\simeq$\,0 is object \#7. {\it (Bottom)} \op/\opp\/ \vs~\sulp/\spp diagram for various \hii\/ region samples: NGC~300 (dots with error bars), M101 (blue open squares), M51 (orange open squares) and additional galaxies taken from \citet[green open triangles]{Bresolin:2005}. In all cases the ionic abundances were derived from the measurement of auroral lines. Two lines of constant $\eta$ have been drawn at $\log\eta$\,=\,0
and $\log\eta$\,=\,0.7. We identify two of the \hii\/ regions discussed in the text, NGC~300-7 and NGC~5471 in M101.\label{fig_eta1}}
\epsscale{1}
\end{figure}

\subsection{Ionizing radiation}\label{ionizing}
As mentioned earlier, the $\eta$\,=\,(\op/\opp)/(\sulp/\spp) parameter introduced by \citet{Vilchez:1988a}
is a measure of the softness of the ionizing radiation, its value decreasing with increasingly harder stellar ionizing continua ($\log\eta\sim1/$\teff\/ in the blackbody case). Fig.~\ref{fig_eta1} {\it (top)} shows a mild radial gradient of $\eta$ in NGC~300, as derived from our ionic aundances. The outlier with a much higher value of $\log\eta$\,$\simeq$\,0 than the rest of the sample is our target \#7 (=\,De\,30), which was identified in \S~\ref{wrsect} as a high-excitation \hii\/ region hosting a WNL star. This object appears as a high-surface brightness nebula, with a compact, round morphology in our H$\alpha$ images.

To illustrate the behavior of $\eta$ in a wider context, we have plotted in Fig.~\ref{fig_eta1} {\it (bottom)}
a \op/\opp\/ \vs~\sulp/\spp diagram that includes the NGC~300 data, as well as other \hii\/ region samples analyzed by our group, for which the ionic abundances were derived by means of auroral lines: M101 (\citealt{Kennicutt:2003}, blue squares), M51 (\citealt{Bresolin:2004}, orange squares), and additional galaxies from \citet[green triangles]{Bresolin:2005}. In this diagram, objects whose nebular spectra can be characterized by similar values of the `effective' temperature of the ionizing radiation field  lie along lines of constant $\eta$, as empirically verified by spatially resolved spectroscopy of nearby \hii\/ regions 
(\citealt{Vilchez:1988a}; \citealt{Kennicutt:2000}).
We have drawn two such lines for $\log\eta$\,=\,0 and $\log\eta$\,=\,0.7 for illustration purposes.
Along lines of constant $\eta$ the ionization parameter $U$, a measure of the ratio of ionizing photon density to the density of atoms, increases towards smaller \sulp/\spp\/ values
(\citealt{Mathis:1985}). 

It can be seen from Fig.~\ref{fig_eta1} that the majority of the \hii\/ galaxies cluster around the $\log\eta$\,=\,0.7 line, with a dispersion that is partly due to metallicity. The region NGC300-7 is one of a few with 
very hard ionizing spectra. Among these we find NGC5471-B (identified in Fig.~\ref{fig_eta1}), NGC5471-A and NGC5471-D (located in proximity of NGC300-7 in the plot), and M51-P203 (orange square near NGC5471-B). The unusually high temperature of the ionizing stars in NGC~5471 was already noted by \citet{Mathis:1982}.
Even harder spectra ($\log\eta$ between $-0.2$ and $+0.35$) can be found among \hii\/ galaxies (\citealt{Hagele:2006,Hagele:2008}), which are more extreme examples of star-forming regions.

We have searched for a correlation between the hardness of the radiation, as measured by $\eta$, and the presence of W-R star features in the nebular spectrum. However, many of the softer spectra (high $\eta$) belong to \hii\/ regions containing W-R stars, mostly of types WC and WNL. The harder spectra could be related to the presence of high-ionization WNE stars (as found in NGC300-7), but although we detect \ciii\lin5696 and \heii\lin4686 in M51-P203 (which has one of the hardest spectra among the objects included in Fig.~\ref{fig_eta1}), we do not find evidence for \nv\/ lines. Besides, broad W-R features are not detected in the various components of NGC~5471 in M101, although nebular \heii\lin4686\/ is present (\citealt{Schaerer:1999}, and confirmed by our independent analysis of the spectra presented by \citealt{Kennicutt:2003}), which is indicative of high-excitation conditions. However, high-excitation is the norm among low-metallicity \hii\/ regions, as is the case of NGC~5471. Still, we think that there is circumstantial evidence to suggest that the hard spectrum of NGC~300-7 is related to the presence of an early WN star among its ionizing sources. \citet{Kennicutt:2000} measured the quantity 
$\eta\prime$, defined by $\log\eta\prime$\,=\,$\log\eta -0.14/t - 0.16$ (\citealt{Vilchez:1988a}), which depends
mildly on the electron temperature ($t$\,=\,\te/10$^4$), in four \hii\/ Galactic regions ionized by W-R stars of type \mbox{WN3-5}, and found values between $-0.2$ and 0.1 (with the lower values for the earlier WN subtype). With an electron temperature around \te\,=\,10$^4$~K, we would then obtain $\log\eta$\,$\simeq$\,0.1--0.3.
This simple example illustrates the fact that the hardness of the nebular spectrum produced by a single ionizing early-type WN star is comparable to what we observe in \mbox{NGC~300-7} ($\log\eta$\,=\,$-0.03$).

\begin{figure}
\medskip
\center \includegraphics[width=0.47\textwidth]{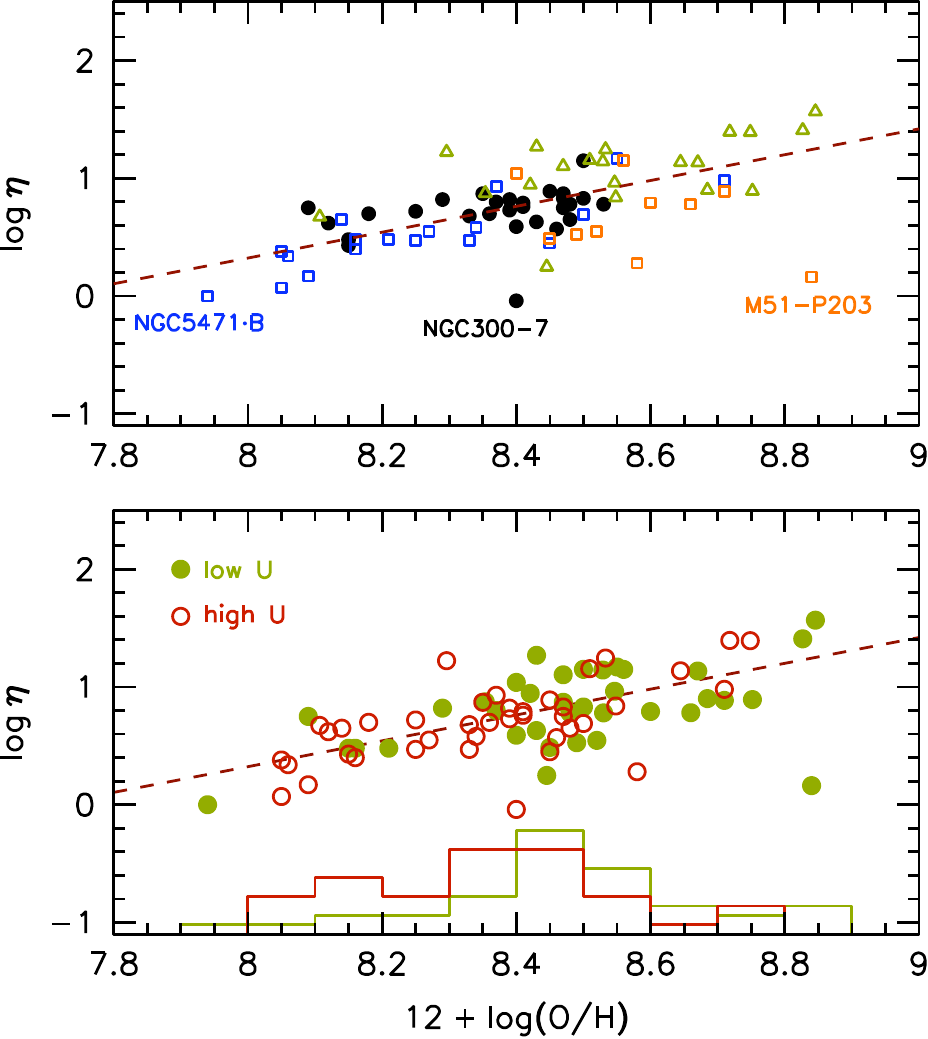}\medskip
\caption{{\it (Top)} The dependence of $\log\eta$ on O/H for the same sample of \hii\/ regions shown
in Fig.~\ref{fig_eta1}. We identify NGC~5471 in M101, NGC~300-7, and M51-P203 as three objects with particularly hard ionizing spectra. The least-square fit to the data points, excluding NGC~300-7 and M51-P203, is shown by the dashed line. {\it (Bottom)} Same diagram as above, using different symbols to separate \hii\/ regions 
with low and high ionization parameter $U$, adopting the median \sulp/\spp\/ ionic ratio as the cut-off value.
The histograms at the bottom, drawn at an arbitrary scale, display the distribution of the two sub-samples in
terms of O/H.\label{fig_eta2}}
\epsscale{1}
\end{figure}

Part of the scatter of the data points in Fig.~\ref{fig_eta1} relative to a line of constant $\eta$ is related to a metallicity trend. This is evident in Fig.~\ref{fig_eta2} {\it (top)}, which shows how $\log\eta$ varies with oxygen abundance for the same sample of \hii\/ regions considered in Fig.~\ref{fig_eta1}.
Excluding NGC~300-7 and M51-P203, the data points follow a linear trend 
defined by the following least-square fit:

\begin{equation}
\log\eta = 1.10~(\pm 0.14)~x~-~8.45~(\pm 1.15)
\end{equation}

\noindent
where $x$\,=\,\oh. This result could, in principle, be affected by the dependence of $\eta$ on the ionization parameter, which is known to vary with metallicity in galaxies (\citealt{Dopita:1986}; \citealt{Bresolin:1999}; \citealt{Dopita:2006b}). We have investigated the effect of the ionization parameter by dividing the \hii\/ region sample into two equal-sized sub-samples, of `high' and `low' $U$, using the fact that the \sulp/\spp\/ ratio, together with its more commonly used observational equivalent \sii/\siii, is a good indicator of this nebular parameter (\citealt{Diaz:1991}). The result is shown in 
Fig.~\ref{fig_eta2} {\it (bottom)}, where the two samples are drawn with different symbols. The histogram at the bottom (drawn on an arbitrary vertical scale) indicates that the distributions of objects in each subsample in terms of O/H are only mildly skewed, 
with  low-$U$ \hii\/ regions favoring high abundances, and high-$U$ \hii\/ regions favoring low abundances.
The effect confirms qualitatively the finding that the ionization parameter is a decreasing function of metallicity.
What is interesting, however, is that in the metallicity range covered by our data the two sub-samples have a similar behavior when we consider the dependence of $\eta$ on O/H. The observed trend does not appear to be generated by the ionization parameter varying systematically with metallicity, as the two sub-samples have approximately the same extent in both O/H and $\log\eta$. If the $\eta$ dependence on $U$ were important, we would expect a strong dichotomy in the diagram, with high-$U$ regions occupying only the low-$\eta$ part of the plot.

The detection of a metallicity trend in the hardness of the ionizing radiation confirms earlier findings by other authors, including \citet{Vilchez:1988a} and \citet{Bresolin:1999}, but using a larger, homogeneous sample of extragalactic \hii\/ regions extending over nearly 0.9 dex in O/H, fully based on \te-based abundances.
The $\eta$ method is successful in ranking stellar temperatures, but the quantification of the observed hardness in terms of an effective temperature of the ionizing sources is model-dependent (\citealt{Mathis:1985}). Assigning \teff\/ values to the $\eta$ scale would require photoionization modeling of \hii\/ regions 
adopting the most recent generation of stellar atmosphere models for hot stars, but this is outside of the scope 
of the present paper. We simply note that, while in some of the older work on the $\eta$ parameter changes in the stellar initial mass function were invoked to explain the softer ionizing field at high metallicity, for example requiring upper mass cut-offs decreasing with metallicity, the current spectral modeling of hot stars at various metallicities provides a natural explanation of the nebular observations. 
Softer ionizing stellar continua  with increasing metallicity are theoretically expected for hot stars, due to the enhanced metal line blocking (\citealt{Kudritzki:2002}; \citealt{Mokiem:2004}). 
\citet{Massey:2004,Massey:2005}, comparing spectra of O stars located in the Milky Way and the Magellanic Clouds, and analyzed with NLTE, line-blanketed model atmospheres, find that early O stars in the SMC are 3000-4000~K hotter than their counterparts in the Milky Way. Population synthesis models
that include line-blanketed atmospheres for O stars and an improved treatment of the ionizing output of W-R stars
are able to reproduce the empirical radiation softening (\citealt{Smith:2002}).


\section{Conclusions and Summary}

The analysis of the emission lines in \hii\/ region spectra has provided the bulk of present-day abundance measurements in nearby galaxies, and is being frequently used to study the chemical composition of high redshift star-forming galaxies. The collection of nebular data is therefore essential to infer the cosmic evolution of metallicity  (\citealt{Savaglio:2005,Maiolino:2008,Lara-Lopez:2009}), although studies of the stellar metallicity
as a function of redshift start to offer an alternative  (\citealt{Panter:2008}).

In the local Universe, the knowledge of abundance gradients in spiral galaxies and their evolution with time
provides the necessary observational constraints to the parameters that drive models of the chemical evolution of galaxies, such as the radial dependence of accretion  and star formation rate in galactic disks (\citealt{Matteucci:1989, Boissier:1999, Chiappini:2001, Colavitti:2008a}).
Excluding the Milky Way and a very few nearby galaxies, such as the Magellanic Clouds, where the metal content of the young stellar populations has been derived from alternative tracers, including Cepheid variables (\citealt{Luck:2003, Romaniello:2008}) and B stars (\citealt{Rolleston:2000, Trundle:2007}), \hii\/ regions often represent the only source of present-day abundance information for star-forming galaxies. 

As explained in the Introduction, nebular abundances are still subject to systematic uncertainties. In the vast majority of cases, extragalactic nebular abundances are derived from strong-line metallicity indices, due to the difficulty of detecting the faint \te-sensitive auroral lines, such as \oiii\lin4363. As discussed by several authors, large differences (up to 0.7 dex) can arise in the derived values, depending on the calibration adopted (\citealt{Kewley:2008}). Empirical calibrations that are tied to auroral line measurements (such as those by \citealt{Pettini:2004}, \citealt{Pilyugin:2005a} and \citealt{Liang:2007a}) provide considerably smaller abundance values (0.2-0.6 dex)
than theoretical calibrations, which are generated from grids of photoionization models (\citealt{McGaugh:1991,Kewley:2002, Tremonti:2004}).
This effect has been strongly manifested after the first direct measurements of electron temperatures in high-metallicity \hii\/ regions (\citealt{Castellanos:2002,Kennicutt:2003,Bresolin:2004, Bresolin:2007}).
The existence of temperature fluctations within nebulae, which could be responsible for the discrepancy between nebular abundances obtained from collisionally excited lines and recombination lines, can, in principle, help to bridge the gap between direct abundances and theoretical ones. However, the importance of temperature fluctuations in explaining the abundance discrepancies in \hii\/ regions is still a controversial topic (\citealt{Stasinska:2007a, Mesa-Delgado:2008}). In light of the importance of \hii\/ region abundances in disparate fields, such as the distance scale (\citealt{Bono:2008}) and the study of gamma-ray burst progenitors (\citealt{Modjaz:2008}),
it is essential to establish the accuracy of the methods used in measuring metallicities.

\medskip
Motivated by these considerations, we have obtained new \hii\/ region spectra in  NGC~300.
In this spiral galaxy about 30 blue supergiants have been analyzed by our group, providing estimates of their metal content, and thus offering the possibility of a comparison between stellar and nebular abundances. What makes 
such a comparison more significant than previous ones is the fact that, instead of having to rely on chemical abundances obtained from strong-line methods (which are subject to the uncertainties mentioned above), we presented a sample of 28 \hii\/ regions
for which we detected auroral lines (\oiii\lin4363, \siii\lin6312, \nii\lin5755) used to derive electron temperatures and O, S, N, Ar and Ne abundances. This procedure is generally considered to provide reliable abundances at least up to the solar metallicity, even though, for the reasons exposed above, it is important to test the results against other methods. The  main conclusion from our work is that our stellar and nebular abundances agree very well, in the range spanned by our objects, \oh\/ approximately between 8.1 and 8.5, i.e.~from the metallicity of the Small Magellanic Clouds to an intermediate value between that of the Large Magellanic Cloud and the Sun.
The central abundance of NGC~300, 
obtained from a linear extrapolation of the \hii\/ region data, is sub-solar, \oh\,=\,8.57\,$\pm$\,0.02, and the 
slope of the radial abundance gradient is $-0.077\pm0.006$ dex\,kpc$^{-1}$. The metallicities of the B and A supergiants analyzed by \citet{Urbaneja:2005a} and \citet{Kudritzki:2008} are fully consistent with the nebular results.
This result leads to our main conclusion, that, in the metallicity range sampled,  {\it direct, \te-based chemical abundances in extragalactic \hiiit\/ \it regions are reliable measures of the nebular abundances.}

The excellent agreement that we find between nebular and stellar abundances does not exclude effects
that would  systematically increase the nebular abundances, such as metal depletion onto dust grains, or temperature fluctuations. However, their combined effect should be less than about 0.13 dex to be compatible 
with the stellar data.
On the other hand, our stellar  abundances do not agree with the case where the \hii\/ region abundances are derived from theoretically calibrated strong-line methods. We showed examples where the latter provide abundances that are 
larger than the stellar ones by 0.3 dex or more in the central galactic region.
The fact that in the case of the Orion nebula the stellar abundances are in good agreement with the gas abundances derived from 
metal recombination lines  does not
imply that a fundamental difference between extragalactic and Galactic \hii\/ regions exists, as far as metallicity determinations are concerned. 
It stresses the need for further comparative studies of stellar and nebular chemical compositions  in the Milky Way and in 
other nearby spiral galaxies, including both collisionally excited lines and metal recombination lines.
It is also worth pointing out that this kind of comparison is somewhat limited by systematic uncertainties, such as those
related to the atomic parameters used to derive the chemical compositions.

We have briefly discussed the detection of W-R star features in our \hii\/ region spectra, and found that in one 
of the nebulae hosting W-R stars the ionizing field has a particularly hard spectrum, as gauged by the $\eta$ parameter. We suggest that this is due to the presence of an early WN star. We have also considered a larger sample
of extragalactic \hii\/ regions where \te-based abundances are available from our previous work, and confirm 
previous findings about a metallicity dependence of $\eta$, in the sense that softer nebular spectra are found 
at higher metallicity.

\acknowledgments
F.B. gratefully acknowledges the support from the National Science Foundation grant AST-0707911.
G.P. and W.G. acknowledge financial support from the Chilean Center for Astrophysics FONDAP 15010003. W.G. also acknowledges support from the BASAL Centro de Astrofisica y Tecnologias Afines for this work.
We thank Grazyna Stasinska for her comments on the manuscript.

\medskip
\noindent
{\it Facilities:} \facility{VLT:Antu (FORS2)}\\

\bibliography{ngc300}

\end{document}